\documentclass[a4paper,twoside]{article}
\newif\ifdraft \global\drafttrue

\usepackage[T1]{fontenc}
\usepackage{a4wide}
\usepackage{times}
\usepackage{graphicx}
\usepackage{theorem}
\usepackage[colorlinks=true,hyperindex=true]{hyperref}
\usepackage{color}
\usepackage{fancyhdr}
\usepackage{amsmath}
\usepackage{bbm}
\usepackage{amsfonts}
\usepackage{enumerate}
\usepackage{cancel}
\ifdraft
\fi
\newtheorem{theorem}{Theorem}[section]
\newtheorem{proposition}[theorem]{Proposition}
\newtheorem{lemma}[theorem]{Lemma}
\newtheorem{definition}[theorem]{Definition}
\newtheorem{corollary}[theorem]{Corollary}
\theoremstyle{plain}
\theorembodyfont{\upshape}

\newcounter{smallarabics}
\newenvironment{arabicenumerate}
{\begin{list}{{\normalfont\textrm{(\arabic{smallarabics})}}}
{\usecounter{smallarabics}\setlength{\itemindent}{0cm}
\setlength{\leftmargin}{5ex}\setlength{\labelwidth}{4ex}
\setlength{\topsep}{0.75\parsep}\setlength{\partopsep}{0ex}
\setlength{\itemsep}{0ex}}}
{\end{list}}

\def\bel{\begin{lemma}}
\def\eel{\end{lemma}}
\def\bec{\begin{corollary}}
\def\eec{\end{corollary}}
\def\bet{\begin{theorem}}
\def\eet{\end{theorem}}
\def\bed{\begin{definition}}
\def\eed{\end{definition}}
\def\bep{\begin{proposition}}
\def\eep{\end{proposition}}
\def\ben{\begin{arabicenumerate}}  
\def\een{\end{arabicenumerate}}
\def\beq{\begin{equation}}
\def\eeq{\end{equation}}  

\def\qed{\hfill$\square$}
\newcommand{\e}{\mathrm{e}}
\let\oldi\i
\newcommand{\myi}{\oldi}
\renewcommand{\i}{\mathrm{i}}

\renewcommand{\d}{\mathrm{d}}

\def\zz{{\mathbb Z}}

\def\rr{{\mathbb R}}
\def\cc{{\mathbb C}}

\def\cL{\mathcal{ L}}

\def\cT{\mathcal{T}}

\def\cK{\mathcal{K}}

\def\cF{\mathcal{F}}

\def\bPt{\bar P^{\kern1pt t}}
\def\bQt{\bar Q^{\kern1pt t}}

\def\fX{\mathfrak{X}}

\newcommand{\rank}{\mathop{\rm rank}\nolimits}

\newcommand{\ie}{{\sl i.e.,\ }}
\newcommand{\eg}{{\sl e.g.,\ }}

\def\Re{\mathrm{Re}\,}

\def\sp{\mathrm{sp}}

\def\tr{\mathrm{tr}}

\def\Ent{{\rm Ent}}

\newcommand{\slim}{\mathop{\mathrm{s\,\textrm{-}\,lim}}\limits}
\def\bar{\overline}
\def\ubar{\underline}
\def\cal{\mathcal}
\renewcommand{\atop}[2]{\genfrac{}{}{0pt}{1}{#1}{#2}}
\begin{document}
\title{\bf\Huge Entropic fluctuations in\\ Gaussian dynamical systems\\ \vskip1cm}
\author{\sc V. Jak\v{s}i\'c$^{a}$,  C.-A. Pillet$^{b}$, A. Shirikyan$^{c}$\\ 
\\ \\ \\ 
$^a$
Department of Mathematics and Statistics, McGill University\\
805 Sherbrooke Street West \\
Montreal,  QC,  H3A 2K6, Canada
 \\ \\
$^b$
Aix-Marseille Universit\'e, CNRS UMR 7332, CPT, 13288 Marseille, France\\
Universit\'e de Toulon, CNRS UMR 7332, CPT, 83957 La Garde, France
\\ \\
$^c$
Department of Mathematics, University of Cergy--Pontoise\\
CNRS UMR 8088, 2 avenue Adolphe Chauvin\\
95302 Cergy--Pontoise, France
}
\def\today{}
\maketitle
\begin{quote}
\noindent{\bf Abstract.} We study non-equilibrium statistical mechanics of a Gaussian dynamical system and compute in closed 
form the large deviation functionals describing the fluctuations of the entropy production observable  with respect to the reference state and 
the non-equilibrium steady state. The entropy production observable of this model is an unbounded function on the phase space, and its 
large deviation functionals  have a surprisingly rich structure. We explore this structure in some detail.
\end{quote}
\thispagestyle{empty}

\tableofcontents
\section{Introduction}
\label{s1}
In this paper, we prove and elaborate  the results announced in Section 9 of~\cite{JPR}. We consider a  dynamical system described by a real separable Hilbert space ${\cal K}$ and the equation of
motion 
\begin{equation}
\frac{\d}{\d t} x_t={\cal L}x_t,\qquad x_0 \in {\cal K},
\label{int-sat}
\end{equation}
where ${\cal L}$ is a bounded linear operator on ${\cal K}$. Let~$D$ be a strictly positive bounded symmetric operator on~${\cal K}$ and $(\fX, \omega_D)$ the Gaussian random field over~${\cal K}$ with zero mean value and covariance~$D$. Eq.~\eqref{int-sat} induces a flow $\phi_{\cal L}=\{\phi_{\cal L}^t\}$ on $\fX$, and our starting point is the dynamical system $(\fX, \phi_{\cal L}, \omega_D)$ (its detailed construction 
is given in Section~\ref{section-grf}). We compute in closed form and  under minimal regularity assumptions  the non-equilibrium characteristics of this model  by exploiting its Gaussian nature. 
In particular, we discuss the existence of a non-equilibrium steady state (NESS), compute the steady 
state entropy production, and study the large deviations of the entropy production observable 
w.r.t.\;both the reference state $\omega_D$ and the NESS. To emphasize the minimal mathematical structure behind 
the results, in the main body of the paper we have adopted  an abstract   axiomatic presentation. 
In Section~\ref{s3}, the results are illustrated on the example of the one-dimensional harmonic crystal.
For additional information and a pedagogical introduction to the theory of entropic fluctuations in  classical 
non-equilibrium statistical mechanics, we refer the reader to the reviews~\cite{RM,JPR}.

There are very few models for which the  
large deviation functionals of the entropy production observable can be computed in  a closed form, and we hope that our results may serve as a guide for  future studies. In addition, an  important characteristic of a Gaussian dynamical system   is  that its  entropy production observable is  an unbounded function on the phase space. This unboundedness has dramatic effects on the form and regularity properties of the large deviation functionals that require modifications of the celebrated fluctuation relations  \cite{ECM, ES, GC1,  GC2}. Although this topic has received a considerable attention in the physics literature \cite{BaCo, BGGZ, BJMS, Fa, HRS, Vi1, Vi2, ZC}, to the best of our knowledge, it has not been studied in the mathematically rigorous literature on the subject.  Thus, another goal of this paper is to initiate a research program dealing with  mathematical theory of extended fluctuation relations in non-equilibrium statistical mechanics, which emerge when some of the usual regularity assumptions (such as compactness of the  phase space, boundedness of the entropy production observable, smoothness of the  time reversal map) are not satisfied.

The paper is organized as follows. In Section \ref{section-grf} we introduce Gaussian dynamical systems. In  Section \ref{sec-epo} we define
the entropy production observable and describe its basic properties. In Section \ref{sec-ness} we introduce the NESS.  Our main results 
are stated in Sections \ref{sec-efref} and \ref{GC}. The entropy production observable is defined as the phase space contraction rate of 
the reference measure $\omega_D$ under the flow $\phi_{\cal L}$, and in Section \ref{sec-perturbations} we examine the effects of a
perturbation of the reference measure on the large deviation theory. In Section \ref{s3} we illustrate our results on two classes of examples, toy models and harmonic chains. 
The proofs are given in Section  \ref{sec-proofs}.

The focus of this paper is the mathematics of the large deviation theory of the entropy production observable. 
The physical implications of our results  will be discussed  in the continuation of this paper \cite{JPS}.

\bigskip
{\noindent\bf Acknowledgment.} 
This research was partly supported by CNRS PICS Project RESSPDE, NSERC (V.J.) and ANR (C.-A.P.\ and A.S.; grants 09-BLAN-0098 and 2011-BS01-015-01).
The work of C.-A.P.\ has been carried out in the framework of the Labex Archim\`ede (ANR-11-LABX-0033) and of the A*MIDEX project (ANR-11-IDEX-0001-02), funded by the ''Investissements d'Avenir'' French Government program managed by the French National Research Agency (ANR).
 C.-A.P.\ and A.S. also acknowledge the hospitality and excellent working conditions at the Mathematics Department of McGill University where part of this work was done. Another part of this work was done during the visits of V.J. to  the Erwin Schr\"odinger Institute in Vienna and 
the  Isaac Newton Institute in Cambridge. V.J. is grateful to these institutions for their hospitality.

\section{The model and results}
\label{s2}
\subsection{Gaussian dynamical systems}
\label{section-grf}  

In order to setup our notation, we  start with some basic facts about classical 
Gaussian dynamical systems. We refer the reader to~\cite{CFS} for a more detailed introduction
to this subject.

Let $\Gamma$ be a countably infinite set and 
\[
\fX=\{x =(x_n)_{n\in \Gamma}\,|\, x_n\in \rr\}=\rr^\Gamma.
\]
For $x\in\fX$ and $I\subset\Gamma$, we denote $x_I=(x_i)_{i\in I}\in\rr^I$. Let $l=(l_n)_{n\in\Gamma}$ be a given sequence of 
strictly positive numbers such that 
$\sum_{n\in\Gamma}l_n=1$ (we shall call such a sequence a {\em weight}).  Then 
\[
d(x, y)=\sum_{n \in \Gamma}l_n \frac{|x_n-y_n|}{1 + |x_n- y_n|}
\]
is a metric on $\fX$ and $(\fX, d)$  is a complete separable metric space. Its Borel $\sigma$-algebra $\cF$ is generated by the set of all cylinders 
$$
C_I(B)= \{ x\in\fX\,|\,x_I \in B \},
$$
where $I\subset\Gamma$ is finite and $B\subset\rr^I$ is a Borel set. 

Let $\nu$ and $\omega$ be two Borel probability measures on $\fX$. We shall write $\nu\ll\omega$
when $\nu$ is absolutely continuous w.r.t.\;$\omega$. The corresponding Radon--Nikodym derivative is denoted by
$$
\Delta_{\nu|\omega}=\frac{\d\nu}{\d\omega}.
$$
We will also use the notation\footnote{Throughout the paper we adopt the convention $\log x=-\infty$ for $x\leq 0$.}
$$
\ell_{\nu|\omega}=\log\Delta_{\nu|\omega}.
$$
The two measures $\nu$ and $\omega$ are called 
equivalent, denoted $\nu\simeq\omega$, if they are mutually absolutely continuous, \ie
$\omega\ll\nu$ and $\nu\ll\omega$. We adopt the shorthand $\nu(f)=\int_\fX f \d \nu$. The relative entropy of $\nu$ w.r.t.\;$\omega$ is defined
as
\beq
\Ent(\nu|\omega)=
\left\{
\begin{array}{ll}
\displaystyle
-\nu(\ell_{\nu|\omega})
&\text{if } \nu\ll\omega,\\[8pt]
-\infty&\text{otherwise}.
\end{array}
\right.
\label{relent}
\eeq
We recall that $\Ent(\nu|\omega)\le0$, with equality iff $\nu=\omega$.
For $\alpha\in\rr$, the relative R\'enyi $\alpha$-entropy of $\nu$ w.r.t.\;$\omega$ is defined as 
$$
\Ent_\alpha(\nu|\omega)=
\left\{
\begin{array}{ll}
\displaystyle
\log\omega\left(\e^{\alpha\ell_{\nu|\omega}}\right)
&\text{if } \nu\ll\omega,\\[8pt]
-\infty&\text{otherwise}.
\end{array}
\right.
$$

We denote by ${\cal K}\subset\fX$ the real Hilbert space with inner product
\beq
(x,y)=\sum_{n\in\Gamma}x_ny_n
\label{twodual}
\eeq
(${\cal K}=\ell_\rr^2(\Gamma)$),  and  by $\{\delta_n\}_{n\in\Gamma}$ its standard basis. The matrix elements of a linear operator~$A$ on~$\ell^2_\rr(\Gamma)$ w.r.t.\;this basis are denoted by  $A_{nm}=(\delta_n, A\delta_m)$.

Let $\fX_l, \fX_{l}^\ast \subset\fX$ be real Hilbert spaces with respective inner products
\[(x,y)_l=\sum_{n\in\Gamma} l_n x_ny_n, \qquad (x,y)_{l^\ast}=\sum_{n\in\Gamma} l_n^{-1} x_ny_n,\] ($\fX_l^\ast$  is the dual of 
$\fX_l$ w.r.t.\;the duality \eqref{twodual}). Clearly, 
$$
\fX_l^\ast\subset\cK\subset\fX_l\subset\fX,
$$
with continuous and dense inclusions. All the measures on $(\fX,\cF)$ we will consider here 
will be concentrated on $\fX_l$.

Let $D$ be a bounded, strictly positive  operator on~$\cK$. The centered Gaussian measure of covariance~$D$ on~$(\fX,\cF)$  is the unique Borel probability measure~$\omega_D$ specified by its value on cylinders
\[
\omega_D(C_I(B)) =\frac1{\sqrt{\det(2\pi D_I)}}
\int_{B}\e^{-\frac{1}{2}(x, D_I^{-1}x)}\d x,
\]
where $D_I= [D_{ij}]_{i,j\in I}$. 
The measure $\omega_D$
is also uniquely specified  by its characteristic function
$$
\fX_l^\ast\ni y\mapsto\chi(y)=\int_\fX\e^{\i(y,x)}\,\d\omega_D(x)=\e^{-(y,Dy)/2}.
$$
The bound 
\beq
\int_\fX\|x\|_{l}^2\d\omega_D(x)=\int_\fX\sum_{n\in \Gamma}l_nx_n^2\,\d\omega_D(x)=\sum_{n\in \Gamma}l_n D_{nn}\le\|D\|,
\label{basicL2}
\eeq
implies that  $\omega_D(\fX\setminus \fX_l)=0$, {\sl i.e.}, that $\omega_D$  is concentrated  on $\fX_l$. 

Let ${\cal T}$ be  the real vector space of all trace class operators on $\cK$ and  $\|T\|_1={\rm tr}((T^\ast T)^{1/2})$ 
the trace norm on ${\cal T}$. The pair~$({\cal T}, \|\cdot\|_{1})$ is a real Banach space. By the 
Feldman--Hajek--Shale theorem, two Gaussian measures~$\omega_{D_1}$ and~$\omega_{D_2}$ on~$(\fX,\cF)$ are equivalent iff $T=D_2^{-1} - D_1^{-1}\in{\cal T}$. In this case, one has
\begin{align}
\Delta_{\omega_{D_2}|\omega_{D_1}}(x)&=\sqrt{\det(I+D_1T)}\,\e^{-(x,Tx)/2},
\label{GaussDeltaForm}\\[2mm]
\Ent(\omega_{D_2}|\omega_{D_1})
&=\frac12\tr\left(D_1T(I+D_1T)^{-1}\right)-\frac12\log\det\left(I+D_1T\right).
\nonumber
\end{align}
Note that $\det\left(I+D_1T\right)=\det\left(I+D_1^{1/2}TD_1^{1/2}\right)
=\det(D_1^{1/2}D_2^{-1}D_1^{1/2})>0$.

Let $\cL$ be  a bounded linear operator on $\cK$ such that
$\cL^\ast\fX_l^\ast\subset\fX_l^\ast$. It follows that $\cL$ has a continuous extension
to $\fX_l$ which we also denote by $\cL$. For $x\in\fX$ and $t\in\rr$ we set 
\begin{equation}
\phi^t_\cL(x)=
\begin{cases}
\e^{t{\cal L}}x &\mbox{if  } x\in\fX_l, \\[3mm]
 x &\mbox{if  } x\not\in\fX_l.
\end{cases}
\label{dyn-gauss}
\end{equation}
The map $(t, x)\mapsto \phi^t_{\cL}(x)$ is measurable and  $\phi_\cL=\{\phi^t_\cL\}_{t\in \rr}$ is a group 
of automorphisms of the measurable space $(\fX,\cF)$ describing the time evolution. We shall call
$\phi_\cL$  the dynamics generated  by $\cL$ and  $(\fX,\phi_\cL,\omega_D)$
a Gaussian dynamical system. Note that for $\omega_D$-almost all $x\in\fX$, $\phi^t_\cL(x)=\e^{t\cL}x$ for all $t\in\rr$.
\subsection{Entropy production observable}
\label{sec-epo}
Our starting point is the dynamical system $(\fX,\phi,\omega)$, where~$\phi$ is the dynamics on~$\fX$ generated by~$\cL$ and~$\omega$  is the centered Gaussian measure with covariance~$D$ (from now on, $\cL$ and $D$ are fixed, and we shall omit explicit reference to them). The measure~$\omega$ is sometimes called the initial or the reference state of the system.
Observables are measurable functions $f:\fX\to\cc$. They evolve according to
$$
f_t(x)=f\circ\phi^t(x).
$$
The expectation of an observable $f$ at time $t\in\rr$ is given by
$$
\omega_t(f)=\omega(f_t)=\int f_t(x)\d\omega(x),
$$
where $\omega_t=\omega\circ\phi^{-t}$ is the centered Gaussian measure
on $(\fX,\cF)$ with covariance
\[
D_t = \e^{t\cL}D\e^{t\cL^\ast}.
\]
$D_t$ is a bounded strictly positive operator on $\ell_\rr^2(\Gamma)$ and 
$\omega_t (\fX_l)=1$ for all $t$. By the Feldman--Hajek--Shale theorem, the two measures 
$\omega_t$ and $\omega$ are equivalent iff $T_t:=D_t^{-1} - D^{-1}\in {\cal T}$. We shall assume more: 
\begin{quote}
\label{G1-def}
{\bf (G1)}  The map $\rr \ni t \mapsto T_t\in {\cal T}$ is differentiable at $t=0$.
\end{quote}
\newcommand{\Gone}{{\hyperref[G1-def]{(G1)}}}

As will be seen later, this condition implies that the function~$t\mapsto T_t$ is differentiable for all $t$. The {\it entropy production observable\/} (or {\it phase space contraction rate\/}) for $(\fX,\phi,\omega)$ is defined by
$$
\sigma(x)=\frac{\d}{\d t}\, \ell_{\omega_t|\omega}(x)\Big|_{t=0}\,, \quad x\in\cK.
$$
A simple computation shows that (cf.~\eqref{ellform})
\begin{equation}
\sigma(x)=(x,\varsigma x) -{\rm tr}(D\varsigma),
\label{ent-gaussian}
\end{equation}
where
\beq
\varsigma =-\frac{1}{2}\frac{\d T_t}{\d t} \Big|_{t=0},\label{varsigmadef}
\eeq
and the derivative is understood in the sense of~$\cT$ (in particular, $\varsigma\in\cT$\,). Since $\mathcal{T}$ is continuously embedded in the Banach space of all bounded operators on~$\cK$, we have
$$
\varsigma =\frac{1}{2}({\cal L^\ast} D^{-1} + D^{-1}{\cal L}).
$$

\noindent{\bf Remark.} If $A$ is a self-adjoint element of $\cT$, then the quadratic form
$(x,Ax)$ has a unique extension from~$\cK$ to an element of 
$L^1(\fX,\d\omega)$. With a slight abuse of notation, we shall also denote this extension by~$(x,Ax)$ (see Lemma~\ref{techno} below
for a more precise statement). Thus, the entropy production observable~\eqref{ent-gaussian} is a continuous function on~$\cK$ and an integrable function on~$\fX$ w.r.t.\ the measure~$\omega$.

\bep\label{gauss-prop-1}
Suppose that {\rm\Gone{}} holds. Then:
\begin{enumerate}[{\rm (1)}]
\item The function $\rr\ni t\mapsto\sigma_t\in L^1(\fX,\d\omega)$ is continuous.
\item $\ell_{\omega_t|\omega}=\int_0^{t}\sigma_{-s}\,\d s$ holds as the Riemann
integral of a continuous $L^1( \fX,\d\omega)$-valued function. It also holds for 
$\omega$-almost every $x\in\fX$ as the Lebesgue integral of a real-valued function.
\item The function $\rr\ni t\mapsto\e^{\ell_{\omega_t|\omega}}\in L^1(\fX,\d\omega)$ is $C^1$ and
\begin{equation} \label{13}
\frac{\d\ }{\d t}\,\e^{\ell_{\omega_t|\omega}}=\e^{\ell_{\omega_t|\omega}}\sigma_{-t}.
\end{equation}

\item $\omega_t(\sigma)=\tr(\varsigma (D_t- D))$ and in particular $\omega(\sigma)=0$.
\item $\Ent(\omega_t|\omega)=-\int_0^t\omega_s(\sigma)\d s$.
\end{enumerate}
\eep

In specific examples, it may happen that only finitely many matrix elements~$\varsigma_{nm}$ are non-zero, and in this case the map~$x\mapsto\sigma(x)$ is continuous on~$\fX$. The function~$\sigma$ is bounded   only  in the trivial case $\sigma=0$. Note that $\sigma=0$ iff $\omega_t=\omega$ for all~$t$; this follows, for instance, from the cocycle property~\eqref{34}. 

\subsection{Non-equilibrium steady state}
\label{sec-ness}
Our next assumptions are:
\begin{quote}
\label{G2-def}
{\bf (G2)} 
There are some numbers $0<m<M<\infty$ such that $m\le D_t\le M$ for all $t\in \rr$.

\label{G3-def}
{\bf (G3)} 
The following strong limits exist:
\[
\slim_{t \rightarrow\pm\infty}D_t = D_\pm.
\]
\end{quote}
\newcommand{\Gtwo}{{\hyperref[G2-def]{(G2)}}}
\newcommand{\Gthree}{{\hyperref[G3-def]{(G3)}}}

It is clear that $m\le D_\pm\le M$, and $\cL D_\pm+D_\pm\cL^\ast=0$.
In what follows, we set 
\begin{equation}
\delta=\frac{m}{M-m}.
\label{def-delta}
\end{equation}
Let $\omega_\pm$ be the centered Gaussian measure
on $(\fX, {\cal F})$ with covariance $D_\pm$.  
\bep\label{gauss-prop-2}
Suppose that {\rm\Gone--\Gthree} hold. Then: 
\begin{enumerate}[{\rm (1)}]
\item For any bounded continuous function $f:\fX\to\rr$,
\[
\lim_{t\rightarrow\pm\infty}\omega_t(f)=\omega_\pm(f).
\]
\item $\sigma \in L^1(\fX, \d\omega_\pm)$ and 
\[
\omega_\pm(\sigma) =\lim_{t\rightarrow \pm\infty}\omega_t(\sigma) 
= \tr(\varsigma (D_\pm-D)).
\]
\end{enumerate}
\eep
Note that 
\[\omega_+(\sigma)=\lim_{t\rightarrow \infty}\frac{1}{t}\int_0^t\omega_s(\sigma)\d s=-\lim_{t\rightarrow \infty}\frac{1}{t}
\Ent(\omega_t|\omega).\]

We shall call $\omega_+$ the NESS and the non-negative number $\omega_+(\sigma)$ 
the entropy production of $(\fX,\phi,\omega)$. 
\subsection{Entropic fluctuations with respect to the reference state}
\label{sec-efref}
Time reversal invariance plays an important role in non-equilibrium statistical mechanics, and in particular in formulation of 
the fluctuation relations. Hence, we shall also
consider the following hypothesis:
\begin{quote}
\label{G4-def}
{\bf (G4)} There exists a unitary involution $\vartheta:\cK\to\cK$ such that  $\vartheta(\fX_l)\subset\fX_l$, 
$\vartheta{\cal L}= -{\cal L}\vartheta$, and $\vartheta D=D \vartheta$. 
\end{quote}
\newcommand{\Gfour}{{\hyperref[G4-def]{(G4)}}}

This assumption implies that $D_{-t}= \vartheta D_t \vartheta$ for all $t\in\rr$, and
thus  $D_-= \vartheta D_+\vartheta$ and $\omega_+=\omega_-\circ\vartheta$. 
Moreover, it follows from Definition~\eqref{varsigmadef} that
$\vartheta\varsigma = -\varsigma\vartheta$. This in turn implies that
$\tr(D\varsigma)=0$ and 
\begin{equation} \label{15}
\sigma(x)=(x,\varsigma x), \quad \omega_+(\sigma)=-\omega_-(\sigma). 
\end{equation}
For simplicity of notation and exposition, we shall  state and prove our main results under the time reversal invariance assumption, 
which covers the cases of physical interest. With a minor modifications of the statements and the proofs, most 
of our results hold without this assumption. We leave these generalizations to the interested reader. 

The relative R\'enyi entropy functional, which is defined by
\beq
e_{t}(\alpha)=\Ent_\alpha(\omega_t|\omega)=\log\omega(\e^{\alpha\ell_{\omega_t|\omega}}),
\label{ealphadef}
\eeq
is a priori finite only for $\alpha \in [0,1]$. To describe its properties, we introduce the sets 
\[
J_{t}=\left\{\alpha\in \rr\,|\,D^{-1} +\alpha T_t>0\right\},\quad t\in\rr,
\]
and denote by $\cc_\pm$ the open upper/lower half-plane.

\bep\label{analprop}
Suppose that {\rm\Gone{}--\Gfour{}} hold.  Then: 
\begin{enumerate}[{\rm (1)}]
\item $J_t=(-\delta_t, 1+\delta_t)$ for some $\delta_t\geq \delta$ and $J_{-t}=J_t$.
\item The function $\alpha\mapsto e_t(\alpha)$ is  finite on the interval $J_t$ and is equal to 
$+\infty$ for $\alpha \not\in  {J_t}$. Moreover, 
this function is convex, extends to an analytic function on the cut plane $\cc_+\cup \cc_-\cup J_t$,
and satisfies
\begin{equation} \label{16}
e_t(0)=e_t(1)=0,
\qquad
e_t'(0)\le0,\quad e_t'(1)\ge0.
\end{equation}
In particular, $e_t(\alpha)\le0$ for $\alpha\in[0,1]$ and $e_t(\alpha)\ge0$ otherwise. 
\item The finite time Evans--Searles symmetry $e_{t}(\alpha) =e_t(1-\alpha)$ holds for all $t$ and $\alpha$.
\end{enumerate}
\eep

We now study the statistical properties of trajectories as $t\to+\infty$. The intervals~$J_t$ do not necessarily form a monotone family, and we define the minimal interval
$$
\ubar J=\liminf_{t\to\infty}J_t=\bigcup_{T>0}\bigcap_{t>T}J_t.
$$
Clearly, one has $\ubar J=(-\ubar\delta, 1+\ubar \delta)$, where $\ubar\delta=\liminf_{t\to\infty}\delta_t\ge\delta$.

\bet\label{EStheorem}
Suppose that {\rm\Gone--\Gfour} hold.
\begin{enumerate}[{\rm (1)}] 
\item 
The limit
\begin{equation}
e(\alpha):=\lim_{t\rightarrow +\infty}\frac{1}{t} e_{t}(\alpha)
\label{grf-es}
\end{equation}
exists for $\alpha \in \ubar J$. Moreover, the function $e(\alpha)$ is convex on the interval $\ubar J$  and satisfies the relations
\begin{equation} \label{SR}
e(0)=e(1)=0,\quad
e^\prime(0)=-\omega_+(\sigma)\le0,\quad
e^\prime(1)=\omega_+(\sigma)\ge0,\quad e(1-\alpha)=e(\alpha).
\end{equation}

\item 
The function $e(\alpha)$ extends to an analytic function on the cut plane $\cc_+\cup\cc_-\cup \, \ubar J$,
and there is a unique signed Borel measure $\nu$ with support contained in
$\rr\setminus \ubar J $ such that $\int|r|^{-1}\d|\nu|(r)<\infty$ and
\begin{equation} \label{20}
e(\alpha)=-\int_\rr\log\left(1-\frac\alpha r\right)\d\nu(r).
\end{equation}

\item 
The Large Deviation Principle holds in the following form. The  function
\[
I(s)= \sup_{-\alpha \in \ubar J}\bigl(\alpha s-e(-\alpha)\bigr)
\]
is convex, takes values in $[0,\infty]$, vanishes only at $s=\omega_+(\sigma)$, and satisfies the Evans--Searles symmetry relation
\begin{equation} \label{ES-relation}
I(-s)=I(s)+s\quad\mbox{for $s\in\rr$}.
\end{equation}
Moreover, there is $\varepsilon>0$ such that, for any open set
${\cal J} \subset (-\omega_+(\sigma)-\varepsilon,\omega_+(\sigma)+\varepsilon)$, we have
\begin{equation} \label{ldp}
\lim_{t \rightarrow \infty} 
\frac{1}{t} \log \omega\left(\left\{
 x\in\fX\,\, \bigg |\,\, \frac{1}{t}\int_0^{t} \sigma_s(x)\, \d s\in {\cal J}\right\}\right) = 
-\inf_{s\in {\cal J}} I(s).
\end{equation}

\item 
The Central Limit Theorem holds. That is, for any Borel set $B\subset\rr$, we have
$$
\lim_{t\to\infty}\omega\left(\left\{x\in\fX\,\, \bigg |\,\, 
\frac{1}{\sqrt t}\int_0^t \left(\sigma_{s}(x)-\omega_+(\sigma)\right)\, \d s\in B
\right\}\right)=\int_B\e^{-x^2/2a}\frac{\d x}{\sqrt{2\pi a}},
$$
where $a=e^{\prime\prime}(1)$. 

\item 
The strong law of large numbers holds. That is, for $\omega$-a.e. $x\in\fX$, we have
\beq
\lim_{t\rightarrow \infty}\frac{1}{t}\int_0^{t}\sigma_s(x)\,\d s 
=\omega_+(\sigma).
\label{slln}
\eeq
\end{enumerate}
\eet 

\noindent{\bf Remark 1.} 
In general, the two limiting measures~$\omega_-$ and~$\omega_+$ are distinct. This property is closely related to the strict positivity of entropy production. In fact, it follows from the second relation in~\eqref{15} that if $\omega_-=\omega_+$, then $\omega_+(\sigma)=0$ as well as $\omega_-(\sigma)=0$, while any of these two conditions imply that the function~$e(\alpha)$ vanishes on~$[0,1]$ and, hence, identically in view of analyticity.

\noindent{\bf Remark 2.} The representation  of $e(\alpha)$ as a logarithmic potential of a signed measure is somewhat surprising, 
and its mathematical and physical significance remains to be studied in the future. 
The measure $\nu$ is related to the spectral measure
of the operator $Q$ (see the proof of Theorem~\ref{EStheorem} for more details). 

\bigskip
Now let $\{t_n\}\subset\rr_+$ be a sequence such that $\delta_{t_n}\to\hat\delta$. We define $\hat J=(-\hat\delta, 1+\hat \delta)$. Note that, by Proposition~\ref{analprop} (1), we have $\hat \delta \geq \delta$. In the case when $\hat \delta$ coincides with $\bar\delta=\limsup_{t\to\infty}\delta_t$, we write~$\bar J$ instead of~$\hat J$. 

\bet\label{EStheorem-bis}
Suppose that {\rm\Gone--\Gfour} hold and $\{t_n\}\subset\rr_+$ is a  sequence  satisfying the above hypothesis. 
\begin{enumerate}[{\rm (1)}] 
\item Let $Q=D_-^{1/2}(D_-^{-1}-D_+^{-1})D_-^{1/2}$. Then 
\beq
-\frac1{\bar \delta}\le Q\le\frac1{1+\bar\delta}.
\label{Qspec}
\eeq
Furthermore, since  the  function $g(z)=z^{-1}\log(1-z)$ is analytic in the cut plane $\cc\setminus[1,\infty)$, the operator-valued function
\begin{equation}
E(\alpha)=-\alpha D_-^{1/2}g(\alpha Q)D_-^{1/2},
\label{never}
\end{equation}
is analytic in the cut plane $\cc_+\cup \cc_-\cup \bar{J}$.
\item For  $\alpha \in \hat J$, the following relation holds:
\begin{equation}
\hat e(\alpha):=\lim_{n\rightarrow \infty}\frac{1}{t_n} e_{t_n}(\alpha)=\tr(E(\alpha)\varsigma),
\label{grf-es-bis}
\end{equation}
and if $\alpha\in\rr$ is not in the closure of~$\hat J$, then
\begin{equation} \label{028}
\limsup_{n\rightarrow \infty}\frac{1}{t_n}e_{t_n}(\alpha)=\infty.
\end{equation}
Moreover, the function $\hat e(\alpha)$ is convex on the interval $\hat J$  and satisfies relations~\eqref{SR}. 

\item 
The Large Deviation Principle holds in the following form. The  function
\begin{equation} \label{28}
\hat I(s)= \sup_{-\alpha \in \hat J}(\alpha s-\hat e(-\alpha))
\end{equation}
is convex, takes values in $[0,\infty]$, vanishes only at $s=\omega_+(\sigma)$, and satisfies the Evans--Searles symmetry relation~\eqref{ES-relation}. Moreover, for any open interval ${\cal J} \subset \rr$, we have
\begin{equation} \label{029}
\lim_{n \rightarrow \infty} 
\frac{1}{t_n} \log \omega\left(\left\{
 x\in\fX\,\, \bigg |\,\, \frac{1}{t_n}\int_0^{t_n} \sigma_s(x)\, \d s\in {\cal J}\right\}\right) = 
-\inf_{s\in {\cal J}} \hat I(s).
\end{equation}
\end{enumerate}
\eet 
{\bf Remark 1.} The functions~$\hat e(\alpha)$ constructed in Theorem~\ref{EStheorem-bis} coincide with~$e(\alpha)$ on the minimal interval~$\ubar J$. Moreover, by Part~(2) of Theorem~\ref{EStheorem-bis}, the functions~$\hat e$ constructed for different sequences~$\{t_n\}$ must coincide on the common domain of definition. 

{\bf Remark 2.} If $\bar \delta=\infty$, then ${\hat e}(\alpha)=e(\alpha)=0$ for $\alpha\in\rr$.

{\bf Remark 3.} The local Large Deviation Principle described in Part~(3) of Theorem \ref{EStheorem} is an immediate consequence of the local 
G\"artner-Ellis theorem (see Appendix A.2 in \cite{JOPP}).
The  global Large Deviation Principle described in Part~(3) of Theorem \ref{EStheorem-bis} cannot be deduced from the G\"artner-Ellis theorem. 
Our proof of the LDP exploits heavily the Gaussian structure of the model and is motivated  by Exercise~2.3.24 in~\cite{DZ}, see also 
\cite{BFL, BFR, BD} for related results.

\subsection{Entropic fluctuations with respect to the NESS}
\label{GC}
We now turn to the statistical properties of the dynamics under the limiting measures~$\omega_\pm$. In view of the time-reversal invariance~\Gfour, it suffices to study the case of one of these measures, and we shall restrict ourselves to~$\omega_+$. Let us set (cf.\ Part~(2) of Proposition~\ref{gauss-prop-1})
$$
e_{t+}(\alpha)=\log\omega_+(\e^{-\alpha\ell_{\omega_t|\omega}})=\log \omega_+\bigl(\e^{-\alpha\int_0^t\sigma_{-s}\,\d s}\bigr)
=\log \omega_+\bigl(\e^{-\alpha\int_0^t\sigma_{s}\,\d s}\bigr),
$$
where the last relation follows from the invariance of~$\omega_+$  under the flow $\phi^t$.
Note that, {\sl a priori,}  $e_{t+}(\alpha)$ might not be finite for any $\alpha\not=0$.

\bet\label{GCtheorem}
Suppose that {\rm\Gone--\Gfour} hold. Then: 
\begin{enumerate}[{\rm (1)}]
\item For any $t\in\rr$, the function $\rr\ni\alpha\mapsto e_{t+}(\alpha)\in(-\infty,+\infty]$
is convex.
\item  The set
\begin{equation} \label{032}
J_t^{+}=\left\{\alpha\in\rr\,|\,D_{+}^{-1} -\alpha T_t>0\right\}
\end{equation}
is an open interval containing $(-\delta,\delta)$, and the function~$e_{t+}(\alpha)$ is real analytic 
on~$J_{t}^+$ and takes value~$+\infty$ on its complement. 
\item 
Let ${\ubar J}^+$ be the interior of the set
\[
\liminf_{t\to\infty}J_t^+=
\bigcup_{T>0}\bigcap_{t>T}J_t^{+}.
\]
Then ${\ubar J}^+$ is an open interval containing $(-\delta, \delta)$. Moreover, for  $\alpha \in {\ubar J}^+$, the limit 
\begin{equation}
e_+ (\alpha)=\lim_{t\rightarrow \infty}\frac{1}{t} e_{t+}(\alpha)
\label{proto-grf-gc}
\end{equation}
exists and defines a real-analytic function on~$\ubar J^+$. Finally, if~$\alpha$ is not in the closure of ${\ubar J}^{+}$, then 
\begin{equation} \label{33}
\limsup_{t\rightarrow \infty}\frac{1}{t} e_{t+}(\alpha)=+\infty.
\end{equation}

\item The Large Deviation Principle holds in the following form. The function 
\[ I^{+}(s)=\sup_{-\alpha \in {\ubar J}^+}(\alpha s -e_+(-\alpha))
\]
is convex, takes values in $[0, \infty]$, and vanishes only at $s=\omega_{+}(\sigma)$.
Moreover, there is an open interval\,~${\mathbb I}^+$ containing~$\omega_+ (\sigma)$ such that, for any open set ${\cal J}\subset {\mathbb I}^{+}$,  
\[
\lim_{t \rightarrow \infty} 
\frac{1}{t} \log \omega_+\left(\left\{ x\in\fX\,\,\bigg |\,\,
 \frac{1}{t}\int_0^t \sigma_{s}(x)\,\d s\in {\cal J}\right\}\right) = 
-\inf_{s\in {\cal J}} I^+(s).
\]

\item The Central Limit Theorem holds. That is, for any Borel set $B\subset\rr$,
$$
\lim_{t\to\infty}\omega_+\left(\left\{x\in\fX\,\, \bigg |\,\, 
\frac{1}{\sqrt t}\int_0^t \left(\sigma_{s}(x)-\omega_+(\sigma)\right)\, \d s\in B
\right\}\right)=\int_B\e^{-x^2/2a_+}\frac{\d x}{\sqrt{2\pi a_+}},
$$
where $a_+=e_+^{\prime\prime}(0)$. 

\item The strong law of large numbers holds. That is, for $\omega_+$-a.e. $x\in\fX$, we have
$$
\lim_{n\rightarrow \infty}\frac{1}{t}\int_0^{t}\sigma_{s}(x)\,\d s 
=\omega_+(\sigma).
$$

\item Let $\ubar J$ be as in Theorem \ref{EStheorem}. Then $e_+(\alpha)=e(\alpha)$ for $\alpha \in {\ubar J}^+\cap \ubar J$. Moreover, there is an open interval ${\mathbb J}^{+}\subset {\mathbb I}^+$ such that $I^{+}(s)=I(s)$ for $s\in {\mathbb J}^+$. 
\end{enumerate}
\eet

{\bf Remark.} This theorem is a refinement of Proposition 9.5 in~\cite{JPR}. We point out that parts (1) and (3) of that proposition are 
inaccurately formulated: in part (1), the interval $(-\delta, 1+\delta)$ has to be replaced with $(-\delta, \delta)$, while in 
part (3) the interval $(-\langle \sigma\rangle_+-\varepsilon, \langle \sigma\rangle_+ +\varepsilon)$ has to be replaced with 
 $(\langle \sigma\rangle_+-\varepsilon, \langle \sigma\rangle_+ +\varepsilon)$.

\bigskip
Finally, we have the following analogue of Theorem~\ref{EStheorem-bis} on statistical properties of the dynamics under the limiting measure~$\omega_+$. Let $\{t_n\}\subset\rr_+$ be an arbitrary increasing sequence going to~$+\infty$ such that the intervals $J_{t_n}^+$ defined by~\eqref{032} converge to a limiting interval~$\hat J^+$. 

\bet \label{GCtheorem-bis}
Under the hypotheses of Theorem~\ref{GCtheorem} the following assertions hold.

\begin{enumerate}[{\rm (1)}]
\item 
For  $\alpha \in \hat J^+$, the limit
\begin{equation}
\hat e_+(\alpha):=\lim_{n\rightarrow \infty}\frac{1}{t_n} e_{t_n+}(\alpha)
\end{equation}
exists and defines a real-analytic function on~$\hat J^+$. If $\alpha$ 
does not belong to the closure of $\hat J^+$, then
$$
\limsup_{n\rightarrow \infty}\frac{1}{t_n}e_{t_n+}(\alpha)=\infty.
$$
Moreover, $\hat e_+(\alpha)$ and $\tr(E(\alpha)\varsigma)$ coincide on their common domain of definition. 

\item 
The Large Deviation Principle holds in the following form. The  function
$$
\hat I^+(s)= \sup_{-\alpha \in \hat J^+}(\alpha s-\hat e_+(-\alpha))
$$
is convex, takes values in $[0,\infty]$ and vanishes only at $s=\omega_+(\sigma)$. Moreover, for any open interval ${\cal J} \subset \rr$, we have
$$
\lim_{n \rightarrow \infty} 
\frac{1}{t_n} \log \omega_+\left(\left\{
 x\in\fX\,\, \bigg |\,\, \frac{1}{t_n}\int_0^{t_n} \sigma_s(x)\, \d s\in {\cal J}\right\}\right) = 
-\inf_{s\in {\cal J}} \hat I^+(s).
$$
\end{enumerate}
\eet

The proof of this result is completely similar to that of Theorem~\ref{EStheorem-bis}, and therefore we omit it. 

{\bf Remark.} Unlike in the case of the Evans-Searles symmetry, there is no a priori reason why the limiting intervals 
${\hat J}^+$ should be symmetric around $\alpha=\tfrac12$, 
and indeed in all cases we know where 
${\hat J}^+$ can be computed, this property does not hold. Hence, the relation   $\hat e_+(\alpha)=\hat e_+(1-\alpha)$ may fail since 
one side may be finite and the other infinite, leading to the failure of the Gallavotti-Cohen symmetry ${\hat I}^+(-s)= {\hat I}^+(s) +s$. 
The fact that for unbounded entropy production observables the Gallavotti-Cohen symmetry may fail is known in the 
physics literature \cite{BaCo, BGGZ, BJMS, Fa, HRS, Vi1, Vi2, ZC}. In these works one can also find various prescriptions how 
the entropy production observable  can be modified so that the Gallavotti-Cohen symmetry is restored. We shall discuss this topic 
in the continuation of this paper \cite{JPS}.

\subsection{Perturbations} 
\label{sec-perturbations}
We shall consider the following type of perturbation of the reference state $\omega$. 
Let $P$ be a  bounded selfadjoint operator on ${\cal K}$ such that $D^{-1} + P>0$. To avoid trivialities, 
we assume that $P$ is not the zero operator. Let 
\[ D^P =(D^{-1}+ P)^{-1}
\]
and let $\omega^P$ be the centered Gaussian measure with covariance $D^P$. Obviously, 
\[ D^P_t = (D_t^{-1} + P_t)^{-1},
\]
where $P_t=\e^{-t {\cal L}^\ast}P\e^{-t {\cal L}}$. 
We consider the following two cases, assuming that {\rm\Gone--\Gfour} hold for $D$.

{\bf Case 1.}  $P$ is a non-negative trace class operator such that
 $\vartheta P=P\vartheta$,  and $\slim_{t\rightarrow\pm \infty}P_t=0$.

In this case, $\omega^P$ and $\omega$ are equivalent and  {\rm\Gone--\Gfour}  also hold for $D^P$.  Moreover, using the superscript~$P$ to denote the objects associated with the initial measure~$\omega^P$, we easily check that 
\[
D_\pm^P=D_\pm, \quad E^P(\alpha) = E(\alpha),\quad \varsigma^P =\varsigma +\frac{1}{2}({\cal L}^\ast P + P{\cal L}),
\quad \omega_{+}^P(\sigma^P)=\omega_{+}(\sigma),
\]
where we used~\eqref{never} to derive the second relation. We also see that
the functions $e^P(\alpha)$ and $e(\alpha)$ coincide on $\ubar J\cap\ubar J^P$.
It is possible, however, that ${\ubar J}^{P}\not={\ubar J}$ and 
${\ubar J}^{+P}\not={\ubar J}^+$, and in fact the difference could be quite dramatic. Indeed, let us fix $P$ and consider the perturbation~$\lambda P$ for $\lambda>0$.
Pick a unit vector $\varphi$ such that $P\varphi=e\varphi$ with $e>0$.

We consider first the case of ${\ubar J}^{\lambda P}$. One easily sees that for any 
$\alpha >1$,
\beq
(\varphi,((D^{\lambda P})^{-1} +\alpha T_t^{\lambda P})\varphi)
\le\frac{\alpha}{m}-\lambda\left(
(\alpha-1)e-\alpha(\varphi,P_t\varphi)\right).
\label{EQ-friday21}
\eeq
There exists $t_0$ such that for $t>t_0$, 
$(\alpha-1)e-\alpha(\varphi,P_t\varphi)>(\alpha-1)e/2$. Hence, for
$t>t_0$ and $\lambda>2\alpha/em(\alpha-1)$ the right hand side of~\eqref{EQ-friday21}
is negative which implies that $\alpha>1+\delta_t^{\lambda P}$. Thus
$$
\ubar\delta^{\lambda P}=\liminf_{t\to\infty}\delta_t^{\lambda P}\le\alpha -1
$$
provided $\lambda>2\alpha/em(\alpha-1)$. Letting now $\alpha\downarrow1$ we 
conclude that
\[
\lim_{\lambda \rightarrow \infty}\ubar \delta^{\lambda P}=0, 
\]
and  the intervals $\ubar J^{\lambda P}$ collapse to $[0,1]$ in the limit $\lambda \rightarrow \infty$. 

To deal with the case of ${\ubar J}^{+\lambda P}$, we
set $\psi_{\alpha,t}=\e^{t\cL}\varphi$ for $\alpha>0$ and $\psi_{\alpha,t}=\varphi$
for $\alpha<0$. A simple analysis yields
$$
(\psi_{\alpha,t},((D_+^{\lambda P})^{-1}-\alpha T_t^{\lambda P})\psi_{\alpha,t})
\le\frac{1+|\alpha|}m \|\psi_{\alpha,t}\|^2-\lambda|\alpha|(e-(\varphi,P_t\varphi)).
$$
Repeating the previous argument, one shows that the length of the interval 
${\ubar J}^{+\lambda P}$ goes to zero as $\lambda\to\infty$, so that the intervals 
${\ubar J}^{+\lambda P}$ collapse to $\{0\}$.

\bigskip
{\bf Case 2.}  $P>0$, $\vartheta P=P\vartheta$,   and $P_t=P$ for all $t\in\rr$. 

Hypotheses {\rm\Gone--\Gfour}  again  hold for $D^P$, and we have 
$$
D_{+}^P= (D_{+}^{-1} + P)^{-1}, \quad \varsigma^P=\varsigma,\quad  \sigma^P=\sigma.
$$ 
Replacing $P$ with $\lambda P$, it is easy to see that 
$\delta^{\lambda P}$, defined by (\ref{def-delta}), satisfies $\lim_{\lambda \rightarrow \infty}\delta^{\lambda P}=\infty$. Since 
$(-\delta^{\lambda P}, 1+\delta^{\lambda P})\subset  {\ubar J}^{\lambda P}$ and  $(-\delta^{\lambda P}, \delta^{\lambda P})\subset  {\ubar J}^{+ \lambda P}$, 
we see that the intervals~${\ubar J}^{\lambda P}$ and~${\ubar J}^{+\lambda P}$  extend to the  whole real line in the limit $\lambda \rightarrow \infty$. 

\section{Examples}
\label{s3}
\subsection{Toy model}
\label{sec-toy}
Suppose that the generator ${\cal L}$ satisfies ${\cal L}^\ast=-{\cal L}$, and let $\varphi\in \cK$ be a unit vector such that the spectral measure for ${\cal L}$ and $\varphi$ is purely absolutely continuous.  Let 
$$
D= I + \lambda P_\varphi,
$$
where $P_{\varphi}=(\varphi,\,\cdot\,)\varphi$ and $\lambda >-1$. 
Then $D_t =I + \lambda P_{\varphi_t}$, where $\varphi_t=\e^{t {\cal L}}\varphi$ is a continuous curve of unit vectors converging weakly to zero as $t\to+\infty$. Let $\lambda_\pm=\frac12(|\lambda|\pm\lambda)$ denote the positive/negative part of~$\lambda$. One easily verifies that {\rm\Gone--\Gthree} hold with
$m=1-\lambda_-$, $M=1+\lambda_+$ and $D_\pm=I$, so that
\[
\delta=\left|\frac12+\frac1\lambda\right|-\frac12.
\]
Without loss of generality we may assume that {\rm \Gfour} holds.\footnote{That can be always achieved by replacing 
$\cK$ with $\cK\oplus \cK$, ${\cal L}$ with ${\cal L} \oplus {\cal L}^\ast$, $\varphi$ with $\frac{1}{\sqrt 2}\varphi\oplus \varphi$, and setting 
$\vartheta(\psi_1\oplus\psi_2)=\psi_2\oplus\psi_1$.}
Since $(I+\lambda P_\psi)^{-1}=I-\frac{\lambda}{1+\lambda}P_\psi$ for any unit vector~$\psi$ and any $\lambda\ne-1$, we have
\begin{align*}
D^{-1} +\alpha T_t&=I-\frac{\lambda}{1+\lambda}
\left((1-\alpha)P_\varphi+\alpha P_{\varphi_t}\right),\\
D_+^{-1} -\alpha T_t&=I-\frac{\lambda}{1+\lambda}
\alpha\left(P_\varphi-P_{\varphi_t}\right).
\end{align*}
Using the simple fact that for any two linearly independent unit vectors $\varphi,\psi$ and all
$a,b\in\rr$,
$$
\sp(aP_\varphi+bP_\psi)
=\{0\}\cup\left\{\frac{a+b}2
\pm\sqrt{\left(\frac{a-b}2\right)^2+ab(\psi,\varphi)^2}\right\},
$$
one easily shows that 
$$
\delta_t=\sqrt{\frac14+\frac{1+\lambda}{\lambda^2(1-(\varphi,\varphi_t)^2)}}
-\frac12,\qquad
J_t^+=\left\{\alpha\in\rr\,\Bigg|\,
|\alpha|<\frac{1+\lambda}{|\lambda|\sqrt{1-(\varphi,\varphi_t)^2}}\right\}.
$$
Recalling that $(\varphi,\varphi_t)\rightarrow 0$ as $t\rightarrow +\infty$
 we see that for all~$\lambda>-1$,
$\ubar\delta=\bar\delta=\delta$ and 
$\ubar J^+=(-\delta^+,\delta^+)$ where
$$
\delta^+=\frac{1+\lambda}{|\lambda|}=\left\{
\begin{array}{clc}
\delta&\mbox{for}&\lambda\in(-1,0],\\[4pt]
1+\delta&\mbox{for}& \lambda\in[0,\infty).
\end{array}
\right.
$$
Furthermore, evaluating Relations~\eqref{eta} and~\eqref{EQ-etplus} 
established below, we obtain
\begin{align*}
e_t(\alpha)&=-\tfrac12\log\left(1+\frac{\lambda^2}{1+\lambda}
\alpha(1-\alpha)\left(1-(\varphi,\varphi_t)^2\right)\right),\\
e_{t+}(\alpha)&=-\tfrac12\log\left(1-\frac{\lambda^2}{(1+\lambda)^2}
\alpha^2\left(1-(\varphi,\varphi_t)^2\right)\right).
\end{align*}
It follows that
$$
\lim_{t\to\infty}\frac1t e_t(\alpha)
=\left\{\begin{array}{lcr}
0&\mbox{for}&|\alpha-\tfrac12|<\tfrac12+\delta,\\[4pt]
+\infty&\mbox{for}&|\alpha-\tfrac12|>\tfrac12+\delta,
\end{array}
\right.\qquad
\lim_{t\to\infty}\frac1t e_{t+}(\alpha)
=\left\{\begin{array}{lcr}
0&\mbox{for}&|\alpha|<\delta^+,\\[4pt]
+\infty&\mbox{for}&|\alpha|>\delta^+.
\end{array}
\right.
$$
Finally, one easily compute the Legendre transforms of these limiting 
functions,
$$
I(s)=(\tfrac12+\delta)|s|-\tfrac12 s,\qquad
I^+(s)=\delta^+|s|.
$$
While the first one satisfies the fluctuation relation, \ie $I(s)+\tfrac12 s$
is an even function, the second one does not.
\subsection{One-dimensional crystal}
\label{sec-harcry}
We follow \cite{JOPP} and  consider the simplest example of the 
one-dimensional  harmonic crystal.   If 
$\Lambda\subset \zz$ is the crystal lattice,    then the  phase space and Hamiltonian  of the harmonic crystal are 
\[\rr^{\Lambda}\oplus \rr^\Lambda=\{(p,q)=(\{p_n\}_{n\in\Lambda}, \{q_n\}_{n\in \Lambda})\,|\, p_n, q_n\in \rr\},\] 
\[ H_\Lambda(p,q)= \sum_{n\in \Lambda}\left( \frac{p_n^2}{2} +\frac{q_n^2}{2} + \frac{(q_n-q_{n-1})^2}{2}\right),\]
where we set  $q_k=0$ for $k\not\in \Lambda$ (Dirichlet boundary conditions). 
The Hamilton equation of motions are 
\[
\left(\begin{matrix} \dot p\\ \dot q\end{matrix}\right)
={\cal L}_\Lambda\left(\begin{matrix}  p\\ q\end{matrix}\right),
\]
where 
\[ {\cal L_\Lambda}= \left(\begin{matrix}0&-j_\Lambda\\
1_\Lambda&0\end{matrix}\right), 
\]
${j}_\Lambda$ is the restriction of the finite difference operator
\beq
(jq)_n=3q_n-q_{n+1} - q_{n-1}
\label{EQ-Allj}
\eeq
to $\rr^\Lambda$ with Dirichlet boundary condition and $1_\Lambda$ the 
identity on $\rr^\Lambda$ (which we shall later identify with the projection
$\rr^\zz\to\rr^\Lambda$). Clearly, for all $\Lambda$, 
$j_\Lambda$ is a bounded selfadjoint operator on $\ell_\rr^2(\Lambda)$ 
satisfying $1\le j_\Lambda\leq 5$.

To fit this model into our abstract framework, we set $\Gamma_\Lambda=\Lambda\times\zz_2$,
$\fX_\Lambda=\rr^{\Gamma_\Lambda}=\rr^\Lambda\oplus\rr^\Lambda$ with the
weight sequence $l=(l_{n,i})_{(n,i)\in\Gamma_\Lambda}$, where  
$l_{n,i}=c_\Lambda(1+n^2)^{-1}$ and  $c_\Lambda$ is a normalization constant. 
One easily verifies that
${\cal L}_\Lambda^\ast\fX_{\Lambda l}^\ast\subset\fX_{\Lambda l}^\ast$ and the dynamics of the harmonic crystal is described by the group $\e^{t\cL_\Lambda}$.
Let $h_\Lambda$ be the self-adjoint operator on 
$\cK_\Lambda=\ell_\rr^2(\Lambda)\oplus\ell_\rr^2(\Lambda)$ associated to
the quadratic form $2H_\Lambda$. Energy conservation implies
$\cL_\Lambda^\ast h_\Lambda+h_\Lambda\cL_\Lambda=0$. Equivalently, the operator 
$L_\Lambda$ defined by
$$
L_\Lambda=h_\Lambda^{1/2}\cL_\Lambda h_\Lambda^{-1/2}
=\left(\begin{matrix}0&-j_\Lambda^{1/2}\\
j_\Lambda^{1/2}&0\end{matrix}\right), 
$$
is skew-adjoint. Since $1\le h_\Lambda\le 5$, this implies in particular that
the group $\e^{t\cL_\Lambda}$ is uniformly bounded on $\cK_\Lambda$.

Our starting point is harmonic crystal on $\Lambda=\zz$ and in this case we drop the subscript $\Lambda$. For our purposes we will 
view this crystal as consisting of three parts, the left, central, and right, specified by 
\[\Lambda_\ell=(-\infty, -1], \qquad  \Lambda_{c}=\{0\}, \qquad \Lambda_r=[1,\infty).
\]
In what follows we, adopt the shorthands $H_{\Lambda_\ell}= H_\ell$, $h_{\Lambda_\ell}=h_\ell$, $j_{\Lambda_\ell}=j_\ell$,  etc. 
Clearly
$$
\fX=\fX_\ell\oplus\fX_c\oplus\fX_r,\qquad
{\cal K}=\cK_\ell\oplus\cK_c\oplus\cK_r,
$$
where $\cK_s=\ell_\rr^2(\Lambda_s)\oplus\ell_\rr^2(\Lambda_s)$ for 
$s=\ell,c, r$, and  
\[
H= H_0+ V_\ell + V_r,
\]
where
\[
H_0=H_\ell + H_c+ H_r 
\]
and $V_\ell(p,q)= -q_0q_{-1}$, $V_r(p, q)=-q_0q_1$.

The reference state $\omega$  is the centered Gaussian measure with covariance 
\[ D=D_\ell\oplus D_c\oplus D_r,
\]
where 
\[D_s= T_s\left(\begin{matrix}I_s&0\\
0&j_s^{-1}\end{matrix}
\right),\qquad s=\ell, c, r,
\]
$I_s$ is the identity on $\ell_\rr^2(\Lambda_s)$, and $T_s>0$. Thus, initially the left/right part of the crystal are  in thermal equilibrium at temperature $T_{\ell/r}$. The Hamiltonian
$V_{\ell/r}$ couples the left/right part  of the crystal to the oscillator located at the site $n=0$ and this allows for the transfer of the energy/entropy between 
these two parts. The entropic fluctuation theorems for this particular Gaussian dynamical system concern statistics of the energy/entropy 
flow between the left and right parts of the crystal.

Hypothesis  {\rm\Gone{}--\Gfour{}} are easily verified following the arguments of Chapter 1 in the lecture notes \cite{JOPP} and 
one finds that 
$$
\omega_{+}(\sigma)=\kappa\frac{(T_\ell- T_r)^2}{T_\ell T_r},
$$
where $\kappa =(\sqrt{5}-1)/2\pi$, and 
\beq
e(\alpha)=-\kappa\log \left( 1 +\frac{(T_\ell-T_r)^2}{T_\ell T_r}\alpha(1-\alpha)\right).
\label{chain-e}
\eeq
Note that $e(\alpha)$ is finite on the interval  $J_o=\,(-\delta_{o},1+\delta_o)$, 
where
\beq
\delta_o=\frac{\min(T_\ell, T_r)}{|T_\ell- T_r|},
\label{EQ-deltao}
\eeq
and takes the value $+\infty$ outside the interval $J_o$. Note also that
$\delta_o$ can take any value in $(0,\infty)$ for appropriate choices of
$T_\ell,T_r\in(0,\infty)$.
The measure $\nu$ in Part (2) of Theorem~\ref{EStheorem} is 
\[
\nu=\kappa {\mathfrak D}_{-\delta_o} + \kappa {\mathfrak D}_{1+\delta_o},
\]
where ${\mathfrak D}_a$ is the Dirac measure centered at $a$. 

We finish this section with several remarks. 

{\bf Remark 1.} The intervals $\ubar J$, $\ubar J^+$ can be strictly 
smaller then $J_o$. To see this, fix $T_c$, $\delta_o$, $\alpha>1$, and set
$T_r=(1+\delta_o^{-1})T_\ell$ to ensure Relation~\eqref{EQ-deltao}.
Let $\varphi\in\cK$ be such that $(\varphi,h_c\varphi)=1$. One has
$$
(\varphi,(D^{-1}+\alpha T_t)\varphi)=\sum_s\frac1{T_s}\left(
(1-\alpha)(\varphi,h_s\varphi)+\alpha(\varphi_t,h_s\varphi_t)\right),
$$
where $\varphi_t=\e^{-t\cL}\varphi$.
Since the skew-adjoint operator $L$ has purely absolutely continuous spectrum and 
$h_c$ is compact, there exists $t_0>0$ such that
$$
(\varphi_t,h_c\varphi_t)
=(\e^{-tL}h^{1/2}\varphi,h^{-1/2}h_c h^{-1/2}\e^{-tL}h^{1/2}\varphi)
<\frac{\alpha-1}{2\alpha}
$$
for all $t>t_0$. Moreover, since the Hamiltonian flow is uniformly bounded
there exists a constant $C$ such that
$$
\frac1{T_{\ell/r}}\left(
(1-\alpha)(\varphi,h_{\ell/r}\varphi)+\alpha(\varphi_t,h_{\ell/r}\varphi_t)
\right)\le C\frac\alpha{T_{\ell}}.
$$
Summing up, if $T_\ell>4CT_c\alpha/(\alpha-1)$, then
$$
(\varphi,(D^{-1}+\alpha T_t)\varphi)\le
\frac{1-\alpha}{2T_c}+2C\frac{\alpha}{T_\ell}<0,
$$
for all $t>t_0$ and hence $\ubar\delta<\alpha$. Thus, in the limit $T_\ell\to\infty$
the interval $\ubar J$ collapses to $[0,1]$. In a similar way one can show that
in the same limit the interval $\ubar J^{+}$ collapses to $\{0\}$. 
On the other hand, arguing as in the Case 2 of Section 
\ref{sec-perturbations}, one can always take $T_{\ell/r}, T_c \rightarrow 0$ in such a way that in this limit the intervals 
$\ubar J$, $\ubar J^+$ extend to the whole real line.  

{\bf Remark 2.} Somewhat surprisingly, even in the simplest example of the harmonic 
crystal discussed in this section, it appears difficult to effectively estimate the 
location of the intervals $\ubar J$, $\ubar J^+$ outside of the perturbative regimes. 
In particular, the subtleties regarding the location of these sets were overlooked in
Sections~1.11, 1.14 and~1.15 of the lecture notes~\cite{JOPP}. These difficulties
raise many interesting questions and we leave the complete analysis of these aspects
as an open problem.

{\bf Remark 3.} An interesting question is whether one can find $P$ such that for the perturbed reference state $\omega^P$ as defined in Section \ref{sec-perturbations} one has $\ubar J= J_o$. That can be done as follows. Set $\beta_s=1/T_s$,
suppose that $\beta_r >\beta_\ell$ and let 
\[
P=\left(\begin{matrix}(\beta_r-\beta_c)1_c&0\\
0&(\beta_r+2\beta_\ell-3\beta_c)j_c+\beta_\ell v_\ell +\beta_r v_r
\end{matrix}\right), 
\]
where $v_{\ell/r}$ denotes the selfadjoint operator associated with the
quadratic form $2V_{\ell/r}$. One easily checks that 
\[
D^P=(\beta_r h - Xh_\ell^{(N)})^{-1},
\]
where $X=\beta_r-\beta_\ell>0$,
$$
h_\ell^{(N)}=\left(\begin{matrix}
1_{\Lambda_\ell\cup\Lambda_c}&0\\
0&j_\ell^{(N)}
\end{matrix}\right), 
$$
and $j_\ell^{(N)}$ denotes the restriction of 
the operator~\eqref{EQ-Allj} to $\rr^{\Lambda_\ell\cup\Lambda_c}$ with 
Neumann boundary condition. We are concerned with the  interval 
\[
J_t^P=\{\alpha \in \rr\,|\, (D^P)^{-1} + \alpha T_t^P>0\}.
\]
Since
\[ 
(D^P_{t})^{-1} =\beta_r h - X\e^{-t {\cal L}^\ast} h_\ell^{(N)}\e^{-t {\cal L}}
=h^{1/2}\left(\beta_r
-X\e^{tL}h^{-1/2}h_\ell^{(N)} h^{-1/2}\e^{-t L}\right) h^{1/2},
\]
a simple computation gives 
\[
(D^P)^{-1}+\alpha T^P_t
= h^{1/2}\left(\beta_r-(1-\alpha)Xh^{-1/2}h_\ell^{(N)}h^{-1/2}
-\alpha X\e^{t L}h^{-1/2}h_\ell^{(N)}h^{-1/2}\e^{-t L}\right) h^{1/2},
\]
and hence
\[
J^P_t=\{\alpha\in\rr\,|\,\beta_r/X>(1-\alpha)h^{-1/2}h_\ell^{(N)} h^{-1/2}
+\alpha\e^{t L}h^{-1/2}h_\ell^{(N)}h^{-1/2}\e^{-t L}\}.
\]
Since $\beta_r/X=1+\delta_o$ and
$$
0\le h_\ell^{(N)}\le h,
$$
we have that for all $t$, 
\[ (-\delta_o, 1+\delta_o)\,\subset J_t^P.\]
Thus, $\lim_{t\rightarrow \infty}\delta_t^P=\delta_o$ and $\ubar J^P=J_o$. 

{\bf Remark 4.} In contrast to Remark~3, we do not know whether there exists $P$ such that for the perturbed reference state $\omega^P$ one has $\ubar J^{+P}= J_o$.

{\bf Remark 5.} In the equilibrium case  $T_\ell= T_r =T$ we have  $\omega_+(\sigma)=0$, and one may naively expect that $\sigma$ does  not fluctuate with 
respect to $\omega$ and $\omega_+$, i.e., that $e(\alpha)=e_+(\alpha)=0$ for all $\alpha$, and that 
$I(s)= I^+(s)=\infty$ if $s \not=0$. If one also takes $T_c=T$ and the perturbed reference state described in Remark 3, then 
$\sigma=0$,  and the above expectation is obviously correct. On the other hand, for the reference state determined by $D$, 
in  the high-temperature regime $T\rightarrow \infty$, $T_c$ fixed, the interval $\ubar J$ collapses  to 
$[0,1]$ while the interval $\ubar J^{+}$ collapses to $\{0\}$. Hence, in this regime, the rate functions ${\hat I}(s)$ and ${\hat I}^+(s)$ are linear for 
$s\leq 0$ and $s\geq  0$, with the slopes of the linear parts determined by the end points of the finite intervals 
${\hat J}$ and ${\hat J}^+$, and the entropy production observable has non-trivial  fluctuations. 

{\bf Remark 6.} The scattering theory arguments of  \cite{JOPP} that lead to the derivation of the formula 
(\ref{chain-e}) extend to the case of  inhomogeneous one-dimensional harmonic crystal with Hamiltonian
\[H_\Lambda(p,q)= \sum_{n\in \Lambda}\left( \frac{p_n^2}{2} +\frac{\omega_n q_n^2}{2} + \frac{\kappa_n(q_n-q_{n-1})^2}{2}\right),
\]
where  $\omega_n$ and~$\kappa_n$ are positive numbers satisfying 
$$
\label{101}
C^{-1}\le \omega_n,\kappa_n\le C\quad\mbox{for all $n\in \zz$},
$$
and $C\ge1$ is a constant. In this case the operator $j$ is the Jacobi matrix
$$
(jq)_n=(\omega_n+\kappa_n+\kappa_{n+1})q_n-\kappa_{n}q_{n-1} -\kappa_{n+1}q_{n+1}, \quad n\in\zz.
$$
One easily verifies that Hypotheses {\rm \Gone{}}, {\rm \Gtwo{}}, and {\rm \Gfour{}} hold. 
If  $j$ has absolutely continuous spectrum (considered as a self-adjoint  operator on $\ell^2_\cc(\zz)$), 
then   {\rm \Gthree{}}  also holds. Moreover, $\omega_+(\sigma)$ and $e(\alpha)$ can be computed in closed form in terms 
of the scattering data of the pair $(j, j_0)$, where $j_0=j_\ell\oplus j_c\oplus j_r$ (for related computations in the context of open quasi-free quantum 
systems we refer the reader to~\cite{JLP,  JOPP, Lan}). The formulas for $\omega_+(\sigma)$ and $e(\alpha)$ involve the scattering 
matrix of the pair $(j, j_0)$\footnote{In the case of harmonic crystal considered in this section, $j$ is a discrete Laplacian and the absolute values of the entries of the scattering matrix of the pair $(j, j_0)$ are either $0$'s or $1$'s. For this reason the formula (\ref{chain-e}) for $e(\alpha)$ has a particularly simple form.}
and estimating the location of the intervals $\ubar J$, $\ubar J^+$ is difficult. However, the  interesting aspect of the  formula 
for $e(\alpha)$ 
is that it  allows to express the measure $\nu$ in Part~(2) of Theorem~\ref{EStheorem} in terms of the scattering data. The mathematical and 
physical significance of this representation remain to be studied in the future. Finally, the scattering methods can be extended to treat  an arbitrary number
of infinite harmonic reservoirs coupled to a finite harmonic system. The discussion of such extensions is beyond the scope of this paper.

\section{Proofs}
\label{sec-proofs}

\subsection{An auxiliary lemma}
Using the notation and conventions of Section~\ref{section-grf}, we have the following simple result.

\bel\label{techno}
\begin{enumerate}[{\rm (1)}]
\item If $A=A^\ast\in\cT$, then the quadratic form $\ell^2_\rr(\Gamma)\ni x\mapsto q_A(x)=(x,Ax)$
has a unique extension to an element of $L^1(\fX,\d\omega_D)$ with a norm satisfying 
$\|q_A\|_1\le\|D\|\,\|A\|_1$. Moreover,
\beq
\int q_A(x)\,\d\omega_D(x)=\tr(DA).
\label{qAform}
\eeq
\item Let $\rr\ni t\mapsto A_t=A^\ast_t\in\cT$ be differentiable at $t=t_0$ and let~$\dot A_{t_0}$ be its derivative. Then the
map $\rr\ni t\mapsto q_{A_t}\in L^1(\fX,\d\omega_D)$ is differentiable at $t=t_0$ and
$$
\left.\frac{\d\ }{\d t}\,q_{A_t}\right|_{t=t_0}=q_{\dot A_{t_0}}.
$$
\item If $1$ does not belong to the spectrum of $A$, then the function
$\mathcal{T}\ni X\mapsto F(X)=\det(I-X)$ is differentiable at $X=A$ and its derivative is given by
\beq
(\mathrm{D}_AF)(X)=-F(A)\,\tr((I-A)^{-1}X).
\label{DFform}
\eeq
\end{enumerate}
\eel

\noindent{\bf Proof}.

\noindent{\bf Part (1)} By Eq.~\eqref{basicL2}, the function $x\mapsto\Phi_y(x)=(y,x)$
belongs to $L^2(\fX,\d\omega_D)$ for $y\in\fX_l^\ast$. Moreover, Fubini's theorem yields
the estimate
\beq
\|\Phi_y\|_2^2=\sum_{i,j\in\Gamma}y_iy_j\int x_ix_j\,\d\omega_D(x)
=\sum_{i,j\in\Gamma}D_{ij}y_iy_j=(y,Dy)\le\|D\|\,\|y\|^2,
\label{PhiEstim}
\eeq
which implies that the linear map $y\mapsto\Phi_y$ has a unique extension 
$\Phi:\ell^2_\rr(\Gamma)\to L^2(\fX,\d\omega_D)$, such that $\|\Phi\|\le\|D\|^{1/2}$.

A self-adjoint $A\in\cT$ has a spectral representation $A=\sum_ka_k\varphi_k(\varphi_k,\,\cdot\,)$,
where the $a_k$ are the eigenvalues of $A$ and the corresponding eigenvectors $\varphi_k$ form 
an orthonormal basis of $\ell^2_\rr(\Gamma)$. 
It follows that $q_A(x)=\sum_ka_k\Phi_{\varphi_k}(x)^2$ from which
we conclude that $q_A$ extends to an element of $L^1(\fX,\d\omega_D)$ with
$$
\|q_A\|_1\le\sum_k|a_k|\,\|\Phi_{\varphi_k}\|^2_2\le\sum_k|a_k|\,\|D\|=\|D\|\,\|A\|_1.
$$
The last equality in Eq.~\eqref{PhiEstim} yields
$$
\int q_A(x)\,\d\omega_D(x)=\sum_ka_k\,\|\Phi_{\varphi_k}\|_2^2
=\sum_ka_k(\varphi_k,D\varphi_k)=\tr(AD),
$$
which proves Identity \eqref{qAform}.

\noindent{\bf Part (2)} It follows from Part (1) that the linear map
$\cT\ni A\mapsto q_A\in L^1(\fX,\d\omega_D)$ is bounded and hence~$C^1$.

\noindent{\bf Part (3)} Using a well known property of the determinant 
(see Theorem~3.5 in~\cite{Si}), we can write
\begin{align*}
F(A+X)=\det(I-(A+X))
&=\det((I-A)(I-(I-A)^{-1}X)\\
&=\det(I-A)\det(I-(I-A)^{-1}X)\\
&=F(A)\det(I-(I-A)^{-1}X).
\end{align*}
To evaluate the second factor on the right-hand side of this identity, we apply the formula
$$
\det(I+Q)=1+\sum_{k=1}^\infty\tr(Q^{\wedge k}),
$$
where $Q^{\wedge k}$ denotes the $k$-th antisymmetric tensor power of $Q$ (see~\cite{Si}).
Since $\|Q^{\wedge k}\|_1\le(k!)^{-1}\|Q\|_1^k$, one has the estimate
$$
|\det(I+Q)-1-\tr(Q)|\le\e^{\|Q\|_1}-1-\|Q\|_1\le\frac{\e^{\|Q\|_1}}2 \|Q\|_1^2.
$$
It follows that
$$
\det(I-(I-A)^{-1}X)=1-\tr((I-A)^{-1}X)+\mathcal{O}(\|X\|_1^2),
$$
as $X\to0$ in $\mathcal{T}$. Thus, we can conclude that 
$$
F(A+X)-F(A)=-F(A)\,\tr((I-A)^{-1}X)+\mathcal{O}(\|X\|_1^2),
$$
and the result follows.\qed

\subsection{Proof of Proposition \ref{gauss-prop-1}}

\noindent{\bf Part (1)} Up to the constant $\tr(D\varsigma)$ (which is well defined
since $\varsigma\in\cT$), $\sigma$ is given by the quadratic form~$q_\varsigma$
which is in $L^1(\fX,\d\omega)$ by Lemma \ref{techno}~(1). For $x\in\fX_l$,
\ie $\omega$-a.e. $x\in\fX$, one has
$$
\sigma_t(x)-\sigma_s(x)=\frac12\left(x,(\e^{t\cL^\ast}\varsigma\e^{t\cL}-
\e^{s\cL^\ast}\varsigma\e^{s\cL})x\right),
$$
whence, setting $\varsigma_t=\e^{t\cL^\ast}\varsigma\e^{t\cL}$ and applying again Lemma~\ref{techno}~(1), it follows that 
$$
\|\sigma_t-\sigma_s\|_{L^1(\fX,\d\omega)}\le \frac12\|D\|\,\|\varsigma_t-\varsigma_s\|_1. 
$$
Thus, it suffices to show that the 
function $t\mapsto\varsigma_t\in\mathcal{T}$ is continuous. This immediately follows from the norm continuity of the group $\e^{t\cL}$, the fact that
$\varsigma\in\mathcal{T}$, and the well-known trace inequality 
$\|AB\|_1\le\|A\|\,\|B\|_1$. We note, in particular, that
$$
\|\sigma_t\|_{L^1(\fX,\d\omega)}\le \|D\|\,(1+\|\e^{t\cL}\|^2)\,\|\varsigma\|_1\quad\mbox{for $t\in\rr$}.
$$

\noindent{\bf Part (2)} From Eq.~\eqref{GaussDeltaForm}, we deduce that
\beq
\ell_{\omega_t|\omega}=\frac12\log\det(I+DT_t)-\frac12q_{T_t}.
\label{ellform}
\eeq
Now note that $T_t=D_t^{-1}-D^{-1}$ satisfies the cocycle relation 
\begin{equation} \label{34}
T_{t+s}=T_t+\e^{-t\cL^\ast}T_s\e^{-t\cL}.
\end{equation}
It thus follows from Assumption \Gone{} that the function $t\mapsto T_t\in\cT$ is 
everywhere differentiable and that its derivative is given by 
\begin{equation} \label{36}
\dot T_t=-2\varsigma_{-t}.
\end{equation}
Lemma~\ref{techno} (3) and the chain rule imply that the first term on the right-hand 
side of~\eqref{ellform}  is a differentiable function of $t$. Using Eq.~\eqref{DFform}, an 
elementary calculation shows that
$$
\frac12\frac{\d\ }{\d t}\log\det(I+DT_t)\Bigr|_{t=0}=-\tr(D\varsigma).
$$
Applying Lemma \ref{techno} (2) to the second term on the right-hand side of
Eq.~\eqref{ellform}, one further gets
$$
-\frac12\frac{\d\ }{\d t}\, q_{T_t}=q_{\varsigma_{-t}}=q_\varsigma\circ\phi^{-t}.
$$
Summing up, we have shown that
$$
\frac{\d\ }{\d t}\,\ell_{\omega_t|\omega}=\sigma_{-t},\quad t\in\rr.
$$
Since the function $t\mapsto\sigma_{-t}\in L^1(\fX,\d\omega)$ is continuous by 
Lemma \ref{techno} (1), and $\ell_{\omega|\omega}=0$, we can use Riemann's integral to write
\beq
\ell_{\omega_t|\omega}=\int_0^t\sigma_{-s}\,\d s.
\label{elltegral}
\eeq
The fact that, for $\omega$-almost every $x\in\fX$, one has
\beq
\ell_{\omega_t|\omega}(x)=\int_0^t\sigma_{-s}(x)\,\d s,
\label{elltegralx}
\eeq
follows from Theorem 3.4.2 in~\cite{HP}.

\noindent{\bf Part (3)} 
From the cocycle relation
\beq
\ell_{\omega_{t+s}|\omega}
=\ell_{\omega_t|\omega}+\ell_{\omega_s|\omega}\circ\phi^{-t},
\label{ellcocycle}
\eeq
we infer
$$
\xi_s=
\frac1s\left(\e^{\ell_{\omega_{t+s}|\omega}}
-\e^{\ell_{\omega_{t}|\omega}}\right)
-\sigma_{-t}\,\e^{\ell_{\omega_{t}|\omega}}
=\frac1s\left(\e^{\ell_{\omega_{s}|\omega}}-1-s\sigma\right)\circ\phi^{-t}\,
\frac{\d\omega_t}{\d\omega},
$$
and hence
$$
\int_\fX|\xi_s|\,\d\omega
=\frac1{|s|}\int_\fX\left|\e^{\ell_{\omega_{s}|\omega}}-1-s\sigma\right|
\,\d\omega
\le\frac1{|s|}\int_\fX\left|\e^{\ell_{\omega_{s}|\omega}}
-1-\ell_{\omega_s|\omega}\right|\,\d\omega
+\frac1{|s|}\int_\fX\left|\ell_{\omega_s|\omega}-s\sigma\right|\,\d\omega.
$$
To prove that Relation~\eqref{13} holds in $L^1(\fX,\d\omega)$, it suffices
to show that both terms on the right-hand side of this inequality vanish
in the limit $s\to0$.

To estimate the first term we note that the inequality 
$\e^\ell-1-\ell\ge0$ (which holds for $\ell\in\rr$) combined with Eq.~\eqref{qAform} and \eqref{ellform} implies
\begin{align*}
\frac1{|s|}\int_\fX\left|\e^{\ell_{\omega_{s}|\omega}}
-1-\ell_{\omega_s|\omega}\right|\,\d\omega
&=\frac1{|s|}\left(\omega(\e^{\ell_{\omega_s|\omega}})-1
-\int_\fX\ell_{\omega_s|\omega}\,\d\omega\right)\\
&=\frac12\left|\frac1{s}\left(\tr(DT_s)-\log\det(I+DT_s)\right)\right|.
\end{align*}
By Assumption \Gone{}, the map $s\mapsto T_s$ is differentiable in $\cT$
at $s=0$. Since $T_0=0$, we can write
$$
\lim_{s\to0}\frac1{|s|}\int_\fX\left|\e^{\ell_{\omega_{s}|\omega}}
-1-\ell_{\omega_s|\omega}\right|\,\d\omega
=\frac12\left|\frac{\d\ }{\d s}\left(\tr(DT_s)-\log\det(I+DT_s)\right)\Big|_{s=0}\right|.
$$
Using Lemma~\ref{techno}~(3) and the chain rule, we get
$$
\frac{\d\ }{\d s}\left(\tr(DT_s)-\log\det(I+DT_s)\right)\Big|_{s=0}
=\tr(D\dot T_0)-\tr(D\dot T_0)=0.
$$
To deal with the second term, we use Eq.~\eqref{elltegral},  Fubini's theorem and 
Lemma~\ref{techno}~(1) to write
\begin{align*}
\frac1{|s|}\int_\fX\left|\ell_{\omega_s|\omega}-s\sigma\right|\,\d\omega
&=\int_\fX\left|\int_0^1\left(\sigma_{-su}-\sigma\right)\,\d u\right|
\le\int_0^1\int_\fX\left|q_{\varsigma_{-su}-\varsigma}\right|\,\d\omega\,\d u\\
&\le\|D\|\int_0^1\|\varsigma_{-su}-\varsigma\|_1\d u,
\end{align*}
and since the map $s\mapsto\varsigma_s$ is continuous in $\cT$, the dominated
convergence theorem yields
$$
\lim_{s\to0}\int_0^1\|\varsigma_{-su}-\varsigma\|_1\d u=0.
$$

\noindent{\bf Part (4)} Relation~\eqref{ent-gaussian} implies that
$$
\omega_t(\sigma)=\omega(\sigma_t)=\int_\fX q_{\varsigma_{t}}\,\d\omega-\tr(D\varsigma),
$$
and formula \eqref{qAform} yields
$$
\omega_t(\sigma)=\tr(D(\varsigma_{t}-\varsigma))=\tr(\varsigma (D_t- D)).
$$

\noindent{\bf Part (5)} Starting from Definition \eqref{relent} and using the cocycle relation
\eqref{ellcocycle}, we obtain
$$
\Ent(\omega_t|\omega)=-\int_\fX\ell_{\omega_t|\omega}\,\d\omega_t
=\int_\fX\ell_{\omega_{-t}|\omega}\,\d\omega.
$$
Eq.~\eqref{elltegralx} and Fubini's theorem further yield
$$
\Ent(\omega_t|\omega)=
\int_\fX\int_0^{-t}\sigma_{-s}\,\d s\d\omega
=-\int_\fX\int_0^{t}\sigma_{s}\,\d s\d\omega
=-\int_0^t\omega_{s}(\sigma)\,\d s.
$$

\subsection{Proof of Proposition \ref{gauss-prop-2}}

\noindent{\bf Part (1)} We have to show that $\omega_+$, the Gaussian measure
of covariance $D_+$, is the weak limit of the net $\{\omega_t\}_{t>0}$. Since
the cylinders form a convergence determining class for Borel measures on~$\fX$ (see Example~2.4 in~\cite{billingsley1999}),
it suffices to show that $\lim_{t\to\infty}\omega_t(C_I(B))=\omega_+(C_I(B))$ holds for
any finite subset $I\subset\Gamma$ and any Borel set $B\subset\rr^I$.
By Hypotheses \Gtwo--\Gthree, one has $\lim_{t\to\infty}D_{t,I}=D_{+,I}$ and
$$
\e^{-\frac12(x,D_{t,I}^{-1}x)}\le\e^{-\frac{\|x\|^2}{2M}},
$$
for all $x\in\rr^I$. It follows that $\lim_{t\to\infty}D_{t,I}^{-1}=D_{+,I}^{-1}$
as well as $\lim_{t\to\infty}\det(2\pi D_{t,I})=\det(2\pi D_{+,I})$ so that
$$
\lim_{t\to\infty}\frac1{\sqrt{\det(2\pi D_{t,I})}}\int_B\e^{-\frac12(x,D_{t,I}^{-1}x)}\,\d x
=\frac1{\sqrt{\det(2\pi D_{+,I})}}\int_B\e^{-\frac12(x,D_{+,I}^{-1}x)}\,\d x,
$$
holds by the dominated convergence theorem. The same argument applies to $\omega_-$.

\noindent{\bf Part (2)} Follows directly from Lemma~\ref{techno} (1) and Proposition~\ref{gauss-prop-1} (4).

\subsection{Proof of Proposition \ref{analprop}}

\noindent{\bf Part (1)}
Let us note that $\alpha\in J_t$ if and only if
\beq
D^{-1}+\alpha(\e^{-t\cL^*}D^{-1}\e^{-t\cL}-D^{-1})>0.
\label{EQ-Friday2}
\eeq
It follows that $J_t$ is open. For $\theta\in[0,1]$, we can write
$$
D^{-1}+\theta\alpha(\e^{-t\cL^*}D^{-1}\e^{-t\cL}-D^{-1})
=\theta\left(D^{-1}+\alpha(\e^{-t\cL^*}D^{-1}\e^{-t\cL}-D^{-1})\right)
+(1-\theta)D^{-1},
$$
whence $\alpha\in J_t\Rightarrow\theta\alpha\in J_t$ and
we can conclude that $J_t$ is an interval.
Multiplying~\eqref{EQ-Friday2} by~$\vartheta$ from the left and the right and using 
the relations $\vartheta=\vartheta^\ast=\vartheta^{-1}$, we obtain
\begin{equation} \label{30}
D^{-1}+\alpha(\e^{t\cL^*}D^{-1}\e^{t\cL}-D^{-1})>0,
\end{equation}
whence we see that $\alpha\in J_{-t}$. By symmetry, we conclude that $J_{-t}=J_t$. Furthermore, multiplying~\eqref{30} by~$\e^{-t\cL^*}$ and~$\e^{-t\cL}$ from the left and the right, respectively, we obtain
$$
\alpha D^{-1}+(1-\alpha)\e^{-t\cL^*}D^{-1}\e^{-t\cL}>0. 
$$
It follows that $1-\alpha\in J_t$, and by symmetry, we conclude that $\alpha\in J_t$ if and only if $1-\alpha\in J_t$. Thus, $J_t$ is an open interval symmetric around
$\alpha=\frac12$.

\noindent{\bf Part (2)}
For any bounded operator $C>0$ on $\ell^2_\rr(\Gamma)$ and for any $\alpha,t\in\rr$ such that $C^{-1}+\alpha T_t>0$, formulas~\eqref{GaussDeltaForm} and~\eqref{ellform} allow us to write
\beq
\e^{\alpha\ell_{\omega_t|\omega}}\,\d\omega_C
=\sqrt{\frac{\bigl(\det(I+DT_t)\bigr)^\alpha}{\det(I+\alpha CT_t)}}\,\d\omega_{(C^{-1}+\alpha T_t)^{-1}}.
\label{gausspert}
\eeq
By definition $D^{-1}+\alpha T_t>0$ for $\alpha\in(-\delta_t,1+\delta_t)$. Taking 
$C=D$ in~\eqref{gausspert} and integrating over~$\fX$, one easily checks that 
\begin{equation} \label{eta}
e_t(\alpha)=\frac\alpha2\log\det(I+DT_t)-\frac12\log\det(I+\alpha DT_t)
\end{equation}
for all $t\in\rr$ and $\alpha\in(-\delta_t,1+\delta_t)$. The first term on the right-hand side of this 
identity is linear in~$\alpha$ and hence entire analytic.\footnote{We shall see in the proof of Theorem~\ref{EStheorem} that it is in fact identically equal to zero.} The determinant in the second term is also 
an entire function of~$\alpha$, and its logarithm is analytic on the set where the operator $I+\alpha DT_t$ is invertible; see Section~IV.1 in~\cite{GK}. Writing $I+\alpha DT_t=D(D^{-1}+\alpha T_t)$, we see that $I+\alpha DT_t$ is invertible for $\alpha\in J_t$. Furthermore, since 
$$
I+\alpha DT_t=\alpha D^{1/2}(\alpha^{-1}I+D^{1/2}T_tD^{1/2})D^{-1/2},
$$
and the operator $D^{1/2}T_tD^{1/2}$ is self-adjoint, we conclude that $I+\alpha DT_t$ is invertible for $\alpha\in\cc\setminus\rr$. 
Hence, the function $e_t(\alpha)$ is analytic in the cut plane $\cc_+\cup \cc_-\cup J_t$. Its convexity is a well-known property of R\'enyi's relative entropy and follows from H\"older's inequality applied to Eq.~\eqref{ealphadef}, and relations~\eqref{16} are easy to check by a direct computation. 

It remains to prove that $e_t(\alpha)=+\infty$ for $\alpha\notin J_t$. To this end, we first note that the spectrum of $D^{-1}$ is contained in the interval $[M^{-1},m^{-1}]$ and that the operator~$\alpha T_t$ is compact. By the Weyl theorem on essential spectrum, it follows that the intersection of the spectrum of the self-adjoint operator $D^{-1}+\alpha T_t$ with the complement of~$[M^{-1},m^{-1}]$ consists of isolated eigenvalues. Thus, if $\alpha\notin J_t$, then there are finitely many orthonormal vectors~$\{\varphi_j\}$, numbers $\lambda_j\ge0$, and an operator $B\ge cI$ with $c>0$ such that
$$
D^{-1}+\alpha T_t=-\sum_{j=1}^n\lambda_j(\varphi_j,\cdot)\varphi_j+B.
$$
It follows that 
\begin{equation} \label{37}
\omega(\e^{\alpha\ell_{\omega_t|\omega}})=\bigl(\det(I+DT_t)\bigr)^{\alpha/2}\int_{\fX}\exp\biggl\{\frac12\sum_{j=1}^n\lambda_j|(\varphi_j,x)|^2\biggr\}\e^{-(x,Bx)/2}\omega(\d x).
\end{equation}
Since $B-D^{-1}\in\cT$ and $D^{-1}+B>0$, we conclude from~\eqref{GaussDeltaForm} that $\e^{-(x,Bx)/2}\omega(\d x)$ coincides, up to a numerical factor~$C>0$, with a centered Gaussian measure whose covariance operator is equal to $D':=(D^{-1}+B)^{-1}$. Hence, we can rewrite~\eqref{37} in the form
$$
\omega(\e^{\alpha\ell_{\omega_t|\omega}})=C\int_{\fX}\exp\biggl\{\frac12\sum_{j=1}^n\lambda_j|(\varphi_j,x)|^2\biggr\}\omega_{D'}(\d x).
$$
Since the support of~$\omega_{D'}$ coincides with the entire space, this integral is infinite.  

\noindent{\bf Part (3)} Using the cocycle relation \eqref{ellcocycle}, we can write\,\footnote{Note that this computation does not use~\Gfour{}.}
\begin{align*}
e_t(1-\alpha)&=\log\omega(\e^{\ell_{\omega_t|\omega}}\e^{-\alpha\ell_{\omega_t|\omega}})
=\log\omega_t(\e^{-\alpha\ell_{\omega_t|\omega}})\\
&=\log\omega(\e^{-\alpha\ell_{\omega_t|\omega}\circ\phi^t})
=\log\omega(\e^{\alpha\ell_{\omega_{-t}|\omega}})=e_{-t}(\alpha).
\end{align*}
Now note that, by~\Gfour{}, the measure~$\omega$ is invariant under~$\vartheta$, whence we conclude that $\omega_{-t}=\omega_t\circ\vartheta$ and $\ell_{\omega_{t}|\omega}\circ\vartheta=\ell_{\omega_{-t}|\omega}$. It follows that $e_{-t}(\alpha)=e_{t}(\alpha)$. Combining this with the above relation, we obtain the Evans--Searles symmetry.

\subsection{Proof of Theorem \ref{EStheorem}}

\noindent{\bf Part (1)} 
We first prove the existence of limit~\eqref{grf-es}. Let us set
\beq
D_t(\alpha)=((1-\alpha)D^{-1}+\alpha D_t^{-1})^{-1}
\label{EQ-Dalphat}
\eeq
and recall that $e_t(\alpha)$ can be written in the form~\eqref{eta}. Using Relations~\eqref{DFform}, \eqref{36}, Lemma~\ref{techno}~(3) and the chain rule we obtain
\begin{equation}
\frac{\d}{\d t}\log\det(I+\alpha D T_t)
=\tr\bigl((I+\alpha D T_t)^{-1}\alpha D\dot T_t\bigr)=-2\alpha\,\tr\bigl(D_t(\alpha)\varsigma_{-t}\bigr)
=-2\alpha\,\tr\bigl(D_{-t}(1-\alpha)\varsigma\bigr).
\label{61}
\end{equation}
In particular, for $\alpha=1$ the derivative is equal to zero for any $t\in\rr$, whence we conclude that the first term in~\eqref{eta} is identically equal to zero. Let us now fix $\alpha\in\ubar J$ and choose $t_0>0$ so large that $\alpha\in J_{t}$ for $t\ge t_0$. It follows from~\eqref{eta} and~\eqref{61} that
\begin{equation} \label{62}
\frac{1}{t}e_t(\alpha)=\frac1te_{t_0}(\alpha)-\frac{2\alpha}{t}\int_{t_0}^t\tr\bigl(D_{-s}(1-\alpha)\varsigma\bigr)\d s. 
\end{equation}
By Assumption~\Gthree
$$
\label{63}
\slim_{s\to\infty}
D_{-s}(1-\alpha)=D_-(1-\alpha):=\bigl(\alpha D^{-1}+(1-\alpha)D_-^{-1}\bigr)^{-1},
$$
and since $\varsigma$ is trace class, it follows that
$$
\lim_{s\to\infty}\tr\bigl(D_s(1-\alpha)\varsigma\bigr) =\tr\bigl(D_-(1-\alpha)\varsigma\bigr).
$$
Combining this with~\eqref{62}, we conclude that for $\alpha\in\ubar J$,
\begin{equation} \label{65}
\lim_{t\to+\infty}\frac1te_t(\alpha)
=-2\alpha\,\tr\bigl(D_-(1-\alpha)\varsigma\bigr). 
\end{equation}

Once the existence of limit is known, we can easily obtain the required properties of~$e(\alpha)$. 
The convexity of~$e(\alpha)$ and the first and last relations in~\eqref{SR} follow immediately from the corresponding properties of~$e_t(\alpha)$. Furthermore, it follows from~\eqref{elltegral} and the invariance of~$\omega$ under~$\vartheta$ that
$$
e_t'(0)=\int_\fX\ell_{\omega_t|\omega}(x)\,\omega(\d x)=\int_\fX\int_0^t\sigma_{-s}(x)\,\d s\,\omega(\d x)
=-\int_\fX\int_0^t\sigma_{s}(x)\,\d s\,\omega(\d x). 
$$
In view of Part~(2), the limit~$e(\alpha)$ is analytic on its domain of definition. By Theorem~25.7 in~\cite{Rock},
$$
\lim_{t\to\infty}\frac1te_t'(\alpha)=e'(\alpha),
$$
for $\alpha\in\ubar J$. Using Fubini's theorem
and Part~(2) of Proposition~\ref{gauss-prop-2}, we derive
$$
e'(0)
=\lim_{t\to\infty}\frac1te_t'(0)
=-\lim_{t\to\infty}\frac1t\int_{0}^t\omega(\sigma_s)\,\d s
=-\omega_+(\sigma)=-\tr(\varsigma D_+). 
$$
The third relation in~\eqref{SR} now follows from the fourth one.

\noindent{\bf Part (2)} 
The analyticity of~$e(\alpha)$ follows from Relation~\eqref{65}. We now prove~\eqref{20}. 

Let $\mu$ be the spectral measure of $Q$ for the linear functional induced by the trace class operator
$D_-^{1/2}\varsigma D_-^{1/2}$. In other words, $\mu$ is the signed Borel measure such that
\beq
\int f(q)\mu(\d q)=\tr(f(Q)D_-^{1/2}\varsigma D_-^{1/2}),
\label{muDef}
\eeq
for any bounded continuous function $f:\rr\to\cc$.
By Eq.~\eqref{Qspec}, the measure $\mu$ has its support in the
interval $[-\bar\delta^{-1},(1+\bar\delta)^{-1}]$. One easily checks that 
$$
f\mapsto\int f(q^{-1})q^{-1}\,\mu(\d q),
$$
defines a continuous linear functional on the Fr\'echet space $C_0(\rr)$ of compactly supported 
continuous functions $f:\rr\to\cc$. By the Riesz representation theorem (see Chapter~2 in~\cite{Rud}), it follows that there exists a  signed Borel  measure~$\nu$, with support on $(-\infty,-\bar\delta]\cup[1+\bar\delta,\infty)$, such that
\beq
\int f(r)\,\nu(\d r)=\int f(q^{-1})q^{-1}\,\mu(\d q)
\label{nuDef}
\eeq
A standard argument based on the monotone class technique shows that~\eqref{nuDef} remains valid for any bounded measurable function~$f$. Decomposing the measures~$\mu$ and~$\nu$ into their positive and negative parts, we easily deduce from~\eqref{nuDef} that
$$
\int f(r)|\nu|(\d r)=\int f(q^{-1})|q|^{-1}|\mu|(\d q),
$$
for all bounded continuous $f$. In particular, taking $f(r)=\frac1r$ outside 
a small neighborhood of zero and using~\eqref{muDef}, we derive
$$
\int\frac{|\nu|(\d r)}{|r|}=\int|\mu|(\d q)
\le\|D_-^{1/2}\varsigma D_-^{1/2}\|_1<\infty.
$$
Recalling relation~\eqref{grf-es-bis} (which will be established below) and using~\eqref{nuDef} with $f(r)=-\log(1-\alpha r^{-1})$ on the support of~$\nu$, we obtain
\begin{align*}
e(\alpha)&=-\alpha\,\tr\bigl(g(\alpha Q)D_-^{1/2}\varsigma D_-^{1/2}\bigr)=-\int\alpha g(\alpha q)\mu(\d q)\\ 
&=-\int q^{-1}\log(1-\alpha q)\mu(\d q)
=-\int\log(1-\alpha r^{-1})\nu(\d r).
\end{align*}
This relation coincides with~\eqref{20}.

To prove the uniqueness, let $\nu_1$, $\nu_2$ be two signed Borel measures with support in $\rr\setminus \ubar J$, satisfying 
$\int|r|^{-1}|\nu_k|(\d r)<\infty$, $k=1,2$, and such that 
\[
\int\log(1-\alpha r^{-1})\nu_1(\d r)
=\int\log(1-\alpha r^{-1})\nu_2(\d r)
\]
for $\alpha \in \ubar J$. Differentiating, we derive that 
\beq 
\int\frac{\d\nu_1(r)}{r-\alpha}=\int\frac{\d\nu_2(r)}{r-\alpha}
\label{cambridge-1}
\eeq
for $\alpha \in \ubar J$. By analytic continuation~\eqref{cambridge-1} holds 
for all $\alpha\in\cc_+\cup\cc_-$. Since the linear 
span of the set of functions $\{(r-\alpha)^{-1}\,|\, \alpha \in \cc_+\cup \cc_-\}$ is dense in $C_0(\rr)$, (\ref{cambridge-1}) yields that for any 
$f\in C_0(\rr)$, $\int f\d\nu_1=\int f\d \nu_2$. Hence $\nu_1=\nu_2$.

\noindent{\bf Part (3)} 
The fact that $I$ is a convex function taking values in~$[0,+\infty]$ follows immediately from the definition. The relation $e'(0)=\omega_-(\sigma)=-\omega_+(\sigma)$ and the regularity of~$e$ imply that~$I$ vanishes only at $s=\omega_+(\sigma)$. The validity of~\eqref{ES-relation} is a straightforward consequence of the last relation in~\eqref{SR}. Let us prove~\eqref{ldp}.

Consider the following family of random variables $\{\Sigma_t\}_{t\in[0,\infty)}$ defined on the probability space $(\fX,\cF,\omega)$
$$
\Sigma_t=\frac1t\int_0^t\sigma_{s}\,\d s.
$$
By Proposition~\ref{gauss-prop-1}~(2) and the symmetry relations $\omega=\omega\circ\vartheta$ and $\sigma\circ\vartheta=-\sigma$, we have
$$
e_t(\alpha)=\log\omega\bigl(\e^{\alpha\ell_{\omega_t|\omega}}\bigr)
=\log\omega\bigl(\e^{\alpha\int_0^t\sigma_{-s}\,\d s}\bigr)
=\log\omega\bigl(\e^{-\alpha\int_0^t\sigma_{s}\,\d s}\bigr)
=\log\omega\bigl(\e^{-\alpha t\Sigma_t}\bigr),
$$
so that $e_t(-\alpha)$ is the cumulant generating function of the family $\{\Sigma_t\}_{t\in[0,\infty)}$.
Applying a local version of the G\"artner--Ellis theorem (see Theorem~4.65 in~\cite{JOPP}), we conclude that~\eqref{ldp} holds with 
$$
\varepsilon=\min\bigl(-\omega_+(\sigma)-\partial^+e(-\ubar\delta),
-\omega_+(\sigma)+\partial^-e(1+\ubar\delta)\bigr)
=\min\bigl(e'(0)-\partial^+e(-\ubar\delta),
\partial^-e(1+\ubar\delta)-e'(1)\bigr)
$$
where $\partial^\pm e(\alpha)$ denotes the right/left derivative of $e(\alpha)$. The fact that $\varepsilon>0$ follows from the convexity and analyticity of~$e(\alpha)$.

\noindent{\bf Part (4)} 
As was shown above, $e_t(-\alpha)$ is the cumulant generating function of $\{\Sigma_t\}$. Therefore, by Bryc's lemma (see~\cite{Br} or Section~4.8.4 in~\cite{JOPP}), the CLT will be established if we prove that~$e_t(\alpha)$ extends analytically to a disc 
$\mathcal{D}_\varepsilon=\{\alpha\in\cc\,|\, |\alpha|<\varepsilon\}$ and satisfies the estimate
\begin{equation} \label{43}
\sup_{t\ge t_0,\alpha\in\mathcal{D}_\varepsilon}\frac1t|e_t(\alpha)|<\infty,
\end{equation}
for some $t_0>0$.
The analyticity was established in Part~(2) of Proposition~\ref{analprop}.
Using the representation~\eqref{62}, one easily sees that in order
to prove~\eqref{43} it suffices to show that
\beq
\sup_{t\in\rr,|1-\alpha|<\varepsilon}
\|D_t(\alpha)\|<\infty.
\label{EQ-easyone}
\eeq
An elementary analysis shows that Assumption~\Gtwo{} implies the lower bound
\beq
(1-\alpha)D_s^{-1}+\alpha D_t^{-1}\ge\frac2M\frac{M-m}{M+m}
\left(\delta+\tfrac12-|\alpha-\tfrac12|\right),
\label{EQ-LowB}
\eeq
for $t,s\in\rr$ and $\alpha\in[-\delta,1+\delta]$. Since for $z\in\cc$
$$
\Re\left((1-z)D_s^{-1}+z D_t^{-1}\right)=(1-\Re z)D_s^{-1}+\Re z D_t^{-1},
$$
we have the upper bound
\beq
\|\left((1-z)D_s^{-1}+z D_t^{-1}\right)^{-1}\|
\le\frac M2\frac{M+m}{M-m}
\left(\delta+\tfrac12-|\Re z-\tfrac12|\right)^{-1}
\label{EQ-UpB}
\eeq
for $s,t\in\rr$ and $z$ in the strip 
$\{z\in\cc\,|\,\Re z\in(-\delta,1+\delta)\}$. Thus,
the required estimate~\eqref{EQ-easyone} holds provided $\epsilon<\delta$.

\noindent{\bf Part (5)} 
We first note that the differentiability of~$e(\alpha)$ at zero and a local version of Theorems~II.6.3 in~\cite{El} (which holds with identical proof) implies that, for any $\varepsilon>0$ and any integer $n\ge1$,
$$
\omega\left(\left\{x\in\fX\,|\,
\left|\Sigma_n-\omega_+(\sigma)\right|\ge\varepsilon\right\}\right)
\le\e^{-a(\varepsilon)n},
$$
where $a(\varepsilon)>0$ does not depend on~$n$. By Theorems~II.6.4 in~\cite{El}, it follows that
\begin{equation} \label{50}
\lim_{n\to\infty}
\frac1n\int_0^n\sigma_s(x)\,\d s=\omega_+(\sigma)
 \end{equation}
for $\omega$-a.e. $x\in\fX$.
Suppose now we have shown the following inequality for some $r<1$
\begin{equation} \label{51}
\sup_{0\le t\le 1}\,\biggl|\int_n^{n+t}\sigma_s(x)\,\d s\biggr|\le (n+1)^r \quad\mbox{for $n\ge n_0(x)$},
\end{equation}
where $n_0(x)\ge0$ is an integer that is finite for $\omega$-a.e.~$x\in\fX$. 
In this case, we can write 
$$
\biggl|\frac1t\int_0^t\sigma_s(x)\,\d s-\frac1n\int_0^n\sigma_s(x)\,\d s\biggr|
\le \frac1n\biggl|\int_n^{n+\hat t}\sigma_s(x)\,\d s\biggr|+\frac{1}{n^2}\biggl|\int_0^n\sigma_s(x)\,\d s\biggr|. 
$$
where $n$ is the integer part of~$t$ and $\hat t=t-n$. It follows from~\eqref{51} that the first term on the right-hand side goes to zero for a.e.~$x\in\fX$, and the second goes to zero in view of~\eqref{50}. Combining this with~\eqref{50}, we obtain~\eqref{slln}. 
Thus, it remains to establish~\eqref{51}. 

Let us fix an arbitrary $r\in(0,1)$ and denote by~$\xi_n(x)$ the expression on the left-hand side of~\eqref{51}. In view of the first relation in~\eqref{15}, we have
$$
\xi_n(x)=\sup_{0\le t\le 1}\biggl|\int_{n}^{n+t}(\e^{s\cL}x,\varsigma \e^{s\cL}x)\d s\biggr|
=\sup_{0\le t\le 1}\bigl|(x,\varsigma_{n,t}\,x)\bigr|, \quad \varsigma_{n,t}:=\int_{n}^{n+t}\varsigma_s\,\d s.
$$
Suppose we have constructed a sequence $\{B_n\}$ of selfadjoint elements of 
$\cT$ such that, for any $n\ge0$,
\begin{equation} \label{54}
\sup_{0\le t\le 1}\bigl|(x,\varsigma_{n,t}\,x)\bigr|\le (x,B_nx), \qquad \|B_n\|_1\le C,
\end{equation}
where $C>0$ does not depend on~$n$. In this case, introducing the events $A_n=\{x\in\fX\,|\,\xi_n(x)\ge (n+1)^r\}$,
for sufficiently small~$\varepsilon>0$, we can write
\begin{equation} \label{55}
\omega(A_n)\le \e^{-\varepsilon (n+1)^r}\omega(\e^{\varepsilon \xi_n})
\le\e^{-\varepsilon (n+1)^r}\bigl(\det(I-2\varepsilon DB_n)\bigr)^{-1/2}, 
\end{equation}
where we used the fact that the Gaussian measures on~$\fX$ with covariance operators $D_\varepsilon'=(D^{-1}-2\varepsilon B_n)^{-1}$ and~$D$ are equivalent, with the corresponding density given by (see~\eqref{GaussDeltaForm})
$$
\Delta_{D_\varepsilon'|D}(x)
=\bigl(\det(I-2\varepsilon DB_n)\bigr)^{1/2}\e^{\varepsilon (x,B_nx)}. 
$$
In view of the second inequality in~\eqref{54}, the determinant in~\eqref{55} is bounded from below by a positive number not depending on~$n\ge0$ for sufficiently small~$\varepsilon>0$. Thus, the series $\sum_n\omega(A_n)$ converges, and by the Borel--Cantelli lemma, inequality~\eqref{51} holds with an almost surely finite integer~$n_0(x)$. 

We now prove~\eqref{54}. From Assumption~\Gtwo{} we derive
$$
M\ge D_t=\e^{t\cL}D\e^{t\cL^\ast}\ge m\,\e^{t\cL}\e^{t\cL^\ast},
$$
so that the uniform bound
\beq
\|\e^{t\cL}\|\le\left(\frac Mm\right)^{1/2},
\label{EQ-UFlow}
\eeq
holds. Since $\varsigma\in\cT$ is selfadjoint, one has
$|(x,\varsigma x)|\le(x,|\varsigma| x)$ for all $x\in\cK$. Hence
$$
\sup_{0\le t\le 1}|(x,\varsigma_{n,t}x)|
\le\int_n^{n+1}|(\e^{s\cL}x,\varsigma\e^{s\cL}x)|\d s
\le\int_n^{n+1}(\e^{s\cL}x,|\varsigma|\e^{s\cL}x)\d s
=(x,B_n x),
$$
where
$$
B_n=\int_n^{n+1}\e^{s\cL^\ast}|\varsigma|\e^{s\cL}\d s
$$
is a self-adjoint element of $\cT$ such that
$$
\|B_n\|_1\le\frac Mm\,\|\varsigma\|_1.
$$
The proof of Theorem~\ref{EStheorem} is complete. 

\subsection{Proof of Theorem~\ref{EStheorem-bis}}
\label{s4.6}

\noindent{\bf Part (1)}
Let $\{s_n\}$ be an arbitrary sequence converging to~$\bar \delta$. Recall that $D^{-1}+\alpha T_{s_n}>0$ for $\alpha\in J_{s_n}$. 
Multiplying this inequality by~$\e^{s_n\cL/2}$ from the right and by~$\e^{s_n\cL^*/2}$ from the left, we obtain
$$
(1-\alpha)D_{-s_n/2}^{-1}+\alpha D_{s_n/2}^{-1}> 0,
$$
for any $\alpha\in J_{s_n}$.
Invoking Assumptions~\Gtwo-\Gthree, we can pass to the limit in the last
inequality to get 
$$
(1-\alpha)D_-^{-1}+\alpha D_+^{-1}\ge 0,
$$
for any $\alpha\in\bar J$.
Taking $\alpha=1+\bar\delta$ and $\alpha=-\bar\delta$ and performing some simple estimation, we obtain inequality~\eqref{Qspec}. Furthermore, it follows from~\eqref{Qspec} that $\alpha Q<1$ for $\alpha\in (-\bar\delta,1+\bar\delta)$, whence we conclude that the operator function~\eqref{never} is analytic in the cut plane $\cc_+\cup \cc_-\cup(-\bar \delta, 1+\bar \delta)$. 

\noindent{\bf Part (2)} 
We first prove the existence of the limit in~\eqref{grf-es-bis}. To this end, 
we shall apply Vitali's convergence theorem to the sequence of functions
$$
h_n(\alpha)=\frac{1}{t_n}e_{t_n}(\alpha), \quad n\ge1,\quad \alpha\in J_{t_n}.
$$ 
By the very definition of $\hat{\delta}$, for any $\varepsilon>0$ there is 
$N_\varepsilon$ such that, for all $n\ge N_\varepsilon$, the function $h_n$ 
is analytic in the cut plane $\cc_-\cup \cc_+\cup\hat J_\varepsilon$
where
$$
\hat J_\varepsilon=(-\hat \delta+\varepsilon, 1+\hat \delta-\varepsilon)
\subset J_{t_n}.
$$
By the proof of Part~(4) of Theorem~\ref{EStheorem} 
(more precisely Eq.~\eqref{EQ-UpB}), the functions $h_n$ are uniformly
bounded in any disk or radius less than $\delta$ around $\alpha=0$.
By the Cauchy estimate, the same is true of their derivatives $h_n'$.
\begin{figure}
\centering
\includegraphics[scale=0.3]{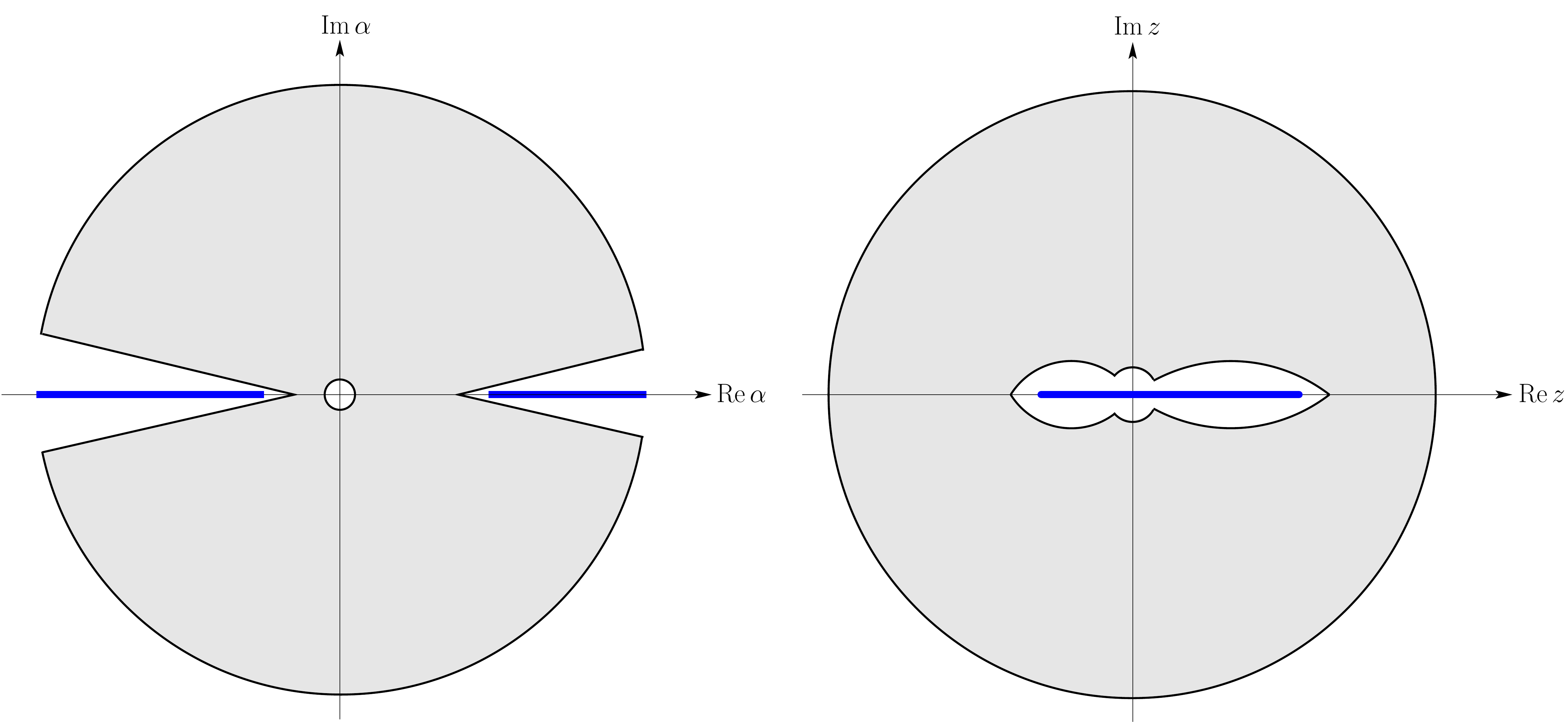}
\caption{A compact region 
$K_0\subset(\cc_-\cup\cc_+\cup\hat J_\varepsilon)\setminus\{0\}$
 and its image under the map
$\alpha\mapsto z=-1/\alpha$. The thick lines in the $\alpha$-plane are the
cuts $\rr\setminus\hat J_\varepsilon$.
 By Eq.~\eqref{EQ-SpecK}, if $n\ge N_\varepsilon$, 
then the spectrum of $Q_n$
lies inside the thick line of the $z$-plane.}
\label{Fig1}
\end{figure}

Let $K_0$ be the compact subset of 
$(\cc_-\cup\cc_+\cup\hat J_\varepsilon)\setminus\{0\}$
described on the left of Figure~\ref{Fig1}.
From Definition~\eqref{EQ-Dalphat} we infer
$$
D_{t_n}(\alpha)=D^{1/2}(1+\alpha Q_n)^{-1}D^{1/2}
=zD^{1/2}(z-Q_n)^{-1}D^{1/2},\quad z=-\frac1\alpha,
$$
where $Q_n=D^{1/2}T_{t_n}D^{1/2}$ is a selfadjoint element of $\cT$.
By definition, $\alpha\in J_{t_n}$ iff $I+\alpha Q_n>0$, \ie 
\beq
\sp(Q_n)\subset(-(1+\delta_{t_n})^{-1},\delta_{t_n}^{-1})
\subset(-(1+\hat\delta-\varepsilon)^{-1},(\hat\delta-\varepsilon)^{-1})
\label{EQ-SpecK}
\eeq
for all $n\ge N_\varepsilon$. Since the function $\alpha\mapsto z=-1/\alpha$
maps $K_0$ to a set which is uniformly separated from $\sp(Q_n)$
(see Figure~\ref{Fig1}), it follows from the spectral theorem that
$$
\sup_{\atop{n\ge N_\varepsilon}{\alpha\in K_0}}\|D_{t_n}(\alpha)\|
\le\|D\|\,\sup_{\atop{n\ge N_\varepsilon}{-z^{-1}\in K_0}}
\frac{|z|}{\mathrm{dist}(z,\sp(Q_n))}<\infty.
$$
Applying Lemma~\ref{techno}~(3) to Eq.~\ref{eta} (recall that the first term
on the right hand side of the latter vanishes) and integrating Eq.~\eqref{36}
to express $T_{t_n}$ we obtain
$$
h_n'(\alpha)=-\frac1{2t_n}\tr(D_{t_n}(\alpha)T_{t_n})
=\int_0^1\tr(D_{t_n}(\alpha)\varsigma_{-st_n})\d s.
$$
The bound~\eqref{EQ-UFlow} further yields
$$
|h_n'(\alpha)|\le\frac Mm\|\varsigma\|_1\,\|D_{t_n}(\alpha)\|,
$$
and the previous estimate allows us to conclude that the sequence
$\{h_n'\}_{n\ge N_\varepsilon}$ is uniformly bounded in $K_0$. 

Summing up, we have shown that $\{h_n'\}_{n\ge N_\varepsilon}$ is uniformly 
bounded on any compact subset of $\cc_-\cup\cc_+\cup\hat J_\varepsilon$ and since $h_n(0)=0$, the same is true of the sequence 
$\{h_n\}_{n\ge N_\varepsilon}$.
By Part~(1) of Theorem~\ref{EStheorem}, the sequence~$\{h_n(\alpha)\}$ 
converges for $\alpha\in\ubar J$. 
By Vitali's theorem (see Section~I.A.12 in~\cite{GR}), we conclude that the 
sequence $\{h_n\}$ converges uniformly on any compact subset of
$\cc_-\cup \cc_+\cup\hat J_\varepsilon$, and the limit is an analytic function 
on it. Since~$\varepsilon>0$ was arbitrary,  we see that the middle term 
in~\eqref{grf-es-bis} is well defined for any 
$\alpha\in\cc_-\cup\cc_+\cup\hat J$ and is an analytic function on this 
domain. 

To prove the second equality in~\eqref{grf-es-bis}, it suffices to establish 
it for $\alpha\in\ubar J$, because both left- and right-hand sides are 
analytic functions on $\cc_-\cup \cc_+\cup\hat J$. The lower 
bound~\eqref{EQ-LowB} shows that $D_t(\alpha)$ is bounded and strictly
positive for all $t\in\rr$ and $\alpha\in(-\delta,1+\delta)$. 
It follows from Eq.~\eqref{ellform} and Lemma \ref{techno} (1) that 
$\ell_{\omega_t|\omega}\in L^1(\fX,\d\omega_{D_t(\alpha)})$. Moreover, 
Eq.~\eqref{gausspert} shows that for $f\in L^1(\fX,\d\omega_{D_t(\alpha)})$
\begin{equation} \label{69}
\omega_{D_t(\alpha)}(f)
=\frac{\omega(\e^{\alpha\ell_{\omega_t|\omega}}f)}
{\omega(\e^{\alpha\ell_{\omega_t|\omega}})}.
\end{equation}
Using this relation with $f=\ell_{\omega_t|\omega}$, integrating the identity
$$
\e^{\alpha\ell_{\omega_t|\omega}}=
1+\int_0^\alpha\e^{\gamma\ell_{\omega_t|\omega}}\ell_{\omega_t|\omega}\,\d\gamma
$$
against~$\omega$, and applying Fubini's theorem, we obtain
$$
\omega(\e^{\alpha\ell_{\omega_t|\omega}})
=1+\int_0^\alpha\omega(\e^{\gamma\ell_{\omega_t|\omega}})
\omega_{D_t(\gamma)}(\ell_{\omega_t|\omega})\,\d\gamma.
$$
Resolving this integral equation (which reduces to a linear differential 
equation) for $\alpha\mapsto\omega(\e^{\alpha\ell_{\omega_t|\omega}})$,
we derive
$$
\omega(\e^{\alpha\ell_{\omega_t|\omega}})=\exp\biggl(\int_0^\alpha \omega_{D_t(\gamma)}(\ell_{\omega_t|\omega})\d\gamma\biggr).
$$
Taking the logarithm, dividing by~$t$, and using~\eqref{elltegralx}, we obtain
\beq
\frac1te_t(\alpha)
=\frac1t\int_0^\alpha\omega_{D_t(\gamma)}(\ell_{\omega_t|\omega})\,\d\gamma
=\frac1t\int_0^\alpha\int_0^t\omega_{D_t(\gamma)}(\sigma_{-s})\,\d s\d\gamma
=\int_0^\alpha\int_0^1\omega_{D_t(\gamma)}(\sigma_{-ts})\,\d s\d\gamma.
\label{e-form}
\eeq
It follows from~\eqref{qAform} and the first relation in~\eqref{15} that
$$
\omega_{D_t(\gamma)}(\sigma_{-ts})=\tr(D_t(\gamma)\,\varsigma_{-ts})
=\tr\bigl(\e^{-ts\cL}D_t(\gamma)\e^{-ts\cL^\ast}\varsigma\bigr)
=\tr\left(\bigl((1-\gamma)D_{-ts}^{-1}+\gamma D_{t(1-s)}^{-1}\bigr)^{-1}\varsigma\right).
$$
Combining this with Hypothesis \Gthree{} and a continuity property of the trace, we derive
$$
\lim_{t\to\infty}\omega_{D_t(\gamma)}(\sigma_{-ts})
=\tr\bigl(\overline{D}_\gamma\varsigma\bigr)=\omega_{\overline{D}_\gamma}(\sigma)\quad
\mbox{for $\gamma\in(-\delta,1+\delta)$, $s\in(0,1)$},
$$
where we set $\overline{D}_\gamma=((1-\gamma)D_-^{-1}+\gamma D_+^{-1})^{-1}$.
The bound~\eqref{EQ-UpB} allows us to apply the dominated convergence theorem
to Eq.~\eqref{e-form}, and  conclude that
\beq
e(\alpha)=\lim_{t\to\infty}\frac1te_t(\alpha)
=\int_0^\alpha\int_0^1\omega_{\overline{D}_\gamma}(\sigma)\,\d s\d\gamma
=\int_0^\alpha\tr\left(\overline{D}_\gamma\varsigma\right)\d\gamma,
\quad \alpha\in(-\delta,1+\delta). 
\label{neckpain}
\eeq
Writing $\overline{D}_\gamma=D_-^{1/2}(I-\gamma Q)^{-1}D_-^{1/2}$, we further get
$$
e(\alpha)
=\int_0^\alpha\tr\bigl(D_-^{1/2}(I-\gamma Q)^{-1}D_-^{1/2}\varsigma\bigr)\d\gamma,
$$
and performing the integral yields Eq.~\eqref{grf-es-bis} for $\alpha\in(-\delta,1+\delta)$.

Finally, to prove~\eqref{028}, it suffices to note that if $\alpha$ does not 
belong to the closure of~$\hat J$ then, for infinitely many $n\ge1$,
$\alpha\notin J_{t_n}$ and by Proposition~\ref{analprop}~(2),
$e_{t_n}(\alpha)=+\infty$.

\noindent{\bf Part (3)} 
The required properties of the rate function~$\hat I$ follow from~\eqref{SR} and elementary properties of the Legendre transform. Thus, we shall only prove~\eqref{029}. In doing so, we shall assume that the interval~$\hat J$ is finite; in the opposite case, the result follows immediately from the G\"artner--Ellis theorem; see Section~4.5.3 in~\cite{DZ}. Moreover, we shall consider only the non-degenerate situation in which $\omega_+(\sigma)>0$. The analysis of the case 
 $\omega_+(\sigma)=0$ is similar and easier.

Let us extend~$\hat e(\alpha)$ to the endpoints of the interval~$\hat J$ by the relation
$$
\hat e(\alpha)=\limsup_{t\to+\infty}\frac1te_t(\alpha), \quad \alpha\in\{-\hat\delta,1+\hat\delta\}. 
$$
Since the extended function~$\hat e$ is convex and, hence, continuous at any point where it is finite, the Legendre transform of~$e(-\alpha)$ coincides with~$\hat I$ defined by~\eqref{28}. In view of a well-known result on the large deviation upper bound (\eg see Theorem~4.5.3 in~\cite{DZ}), the following inequality holds for any closed subset $F\subset\rr$:
$$
\limsup_{n \rightarrow \infty} 
\frac{1}{t_n} \log \omega\left(\left\{
 x\in\fX\,\, \bigg |\,\, \frac{1}{t_n}\int_0^{t_n} \sigma_s(x)\, \d s
 \in F\right\}\right) \le-\inf_{s\in F} \hat I(s).
$$
Since~$\hat I$ is also continuous, this upper bound easily implies that~\eqref{028} will be established if we prove the inequality
\begin{equation} \label{75}
\liminf_{n \rightarrow \infty} 
\frac{1}{t_n} \log \omega\left(\left\{
 x\in\fX\,\, \bigg |\,\, \frac{1}{t_n}\int_0^{t_n} \sigma_s(x)\, \d s
 \in O\right\}\right) \ge-\inf_{s\in O} \hat I(s),
\end{equation}
where $O\subset\rr$ is an arbitrary open set. A standard argument shows that it suffices to prove~\eqref{75} for any open interval ${\cal J}\subset\rr$. Let us set
$$
s^-=-\lim_{\alpha\uparrow1+\hat\delta}\hat e'(\alpha),\qquad
s^+=-\lim_{\alpha\downarrow-\hat\delta}\hat e'(\alpha).
$$
In view of the local version of the G\"artner--Ellis theorem (see Theorem~4.65 in\footnote{In the formulation of Theorem~4.65 in~\cite{JOPP}, it is required that the limit of $t_n^{-1}e_{t_n}(\alpha)$ as $n\to\infty$ should exist for any $\alpha$ in the closure of~$\hat J$. However, the same proof works also in the case when the limits exist  only for $\alpha\in\hat J$.}~\cite{JOPP}), relation~\eqref{029} is true for any interval~${\cal J}\subset (s^-,s^+)$. Thus, it  suffices to consider the case when~$\cal J=\cal J_{s,\varepsilon}=(s-\varepsilon,s+\varepsilon)$, where $\pm (s-s_\pm)\ge0$. The proof of~\eqref{75} is divided into several steps. 

{\it Step~1: Reduction}. We first show that the required inequality will be established if we prove that, for any $\hat s\in\rr$ satisfying the inequality $\pm(\hat s- s_\pm)\ge0$ and any $\varepsilon>0$,  
\begin{equation} \label{76}
\liminf_{n\to\infty}\frac{1}{t_n}\log\omega\bigl(B_n(\hat s,\varepsilon)\bigr)\ge -\hat I(\hat s\pm\varepsilon),
\end{equation}
where $B_n(\hat s,\varepsilon)=\{x\in\fX\,|\,|t_n^{-1}\ell_{\omega_{t_n}|\omega}+\hat s|<\varepsilon\}$. Indeed, we have
\begin{equation} \label{80}
\hat I(s)=
\left\{
\begin{array}{cl}
-(1+\hat \delta)s-e^-&\mbox{ for $s\le s^-$},\\[4pt]
\hat \delta s-e^+&\mbox{ for $s\ge s^+$},
\end{array}
\right.
\end{equation}
where~$e^-$ (respectively, $e^+$) is the limit of~$\hat e(\alpha)$ as $\alpha\uparrow1+\hat\delta$ (respectively, $\alpha\downarrow-\hat\delta$). In particular, the rate function~$\hat I$ is everywhere finite and  continuous. It follows from~\eqref{76} and inequality~\eqref{75} with $\cal J\subset(s^-,s^+)$  that
\begin{align*}
\lim_{\varepsilon\to0^+}\liminf_{n\to\infty}\frac{1}{t_n}\log
\omega\left(\left\{x\in\fX\,\, \bigg |\,\, \frac{1}{t_n}\int_0^{t_n} \sigma_s(x)\, \d s\in {\cal J}_{\hat s,\varepsilon}\right\}\right)
&=\lim_{\varepsilon\to0^+}\liminf_{n\to\infty}\frac{1}{t_n}\log\omega\bigl(B_n(\hat s,\varepsilon)\bigr)\\
&\ge -\hat I(\hat s), 
\end{align*}
where $\hat s\in\rr$ is any point. A well-known (and simple) argument implies the required lower bound~\eqref{75} for any interval~$\cal J\subset\rr$. 
Thus, we need to establish~\eqref{76}. To simplify the notation, we shall consider only the case when $\hat s\ge s_+$ (assuming that $s_+<\infty$).

{\it Step~2: Shifted measures}. 
Let us fix $\hat s\ge s^+$ and denote $\tilde e_t(\alpha)=e_t(-\alpha)$ and $\tilde e(\alpha)=\hat e(-\alpha)$. Since $\tilde e_{t_n}'$ is a monotone increasing function mapping the interval $-J_{t_n}=(-1-\delta_{t_n},\delta_{t_n})$ onto $(-\infty,\infty)$
(see~\eqref{eta}), for any $n\ge1$ there is a unique number $\alpha_n\in- J_{t_n}$ such that $\tilde e_{t_n}'(\alpha_n)=t_n\hat s$. Following a well-known idea in the theory of large deviations, let us define a sequence of measures~$\nu_n$ on~$\fX$ by their densities
$$
\Delta_{\nu_n|\omega}=\exp\bigl(-\alpha_n\ell_{\omega_{t_n}|\omega}-\tilde e_{t_n}(\alpha_n)\bigr).
$$
Suppose we have proved that 
\begin{equation} \label{79}
\liminf_{n\to\infty}\nu_n\bigl(B_n(\hat s,\varepsilon)\bigr)>0.
\end{equation}
In this case, assuming that $\alpha_n>0$, we can write
\begin{align*}
\omega\bigl(B_n(\hat s,\varepsilon)\bigr)
&=\int_{B_n(\hat s,\varepsilon)}
\exp\bigl(\alpha_n\ell_{\omega_{t_n}|\omega}+\tilde e_{t_n}(\alpha_n)\bigr)\d\nu_n\\
&\ge \exp\bigl(t_n\alpha_n(-\hat s-\varepsilon)+\tilde e_{t_n}(\alpha_n)\bigr)
\nu_n\bigl(B_n(\hat s,\varepsilon)\bigr),  
\end{align*}
whence it follows that
\begin{equation} \label{77}
\liminf_{n\to\infty}\frac{1}{t_n}\log\omega\bigl(B_n(\hat s,\varepsilon)\bigr)
\ge \liminf_{n\to\infty}\Bigl(\alpha_n(-\hat s-\varepsilon)+\frac{1}{t_n}\tilde e_{t_n}(\alpha_n)\Bigr). 
\end{equation}
If we know that
\begin{equation} \label{78}
\lim_{n\to\infty}\alpha_n=\hat\delta,\quad 
\liminf_{n\to\infty}\frac{1}{t_n}\tilde e_{t_n}(\alpha_n)\ge e^+,
\end{equation}
then $\alpha_n>0$ for $n$ large enough and inequality~\eqref{77} and relation~\eqref{80} immediately imply the required result~\eqref{76}. 
Thus, we need to prove~\eqref{79} and~\eqref{78}.

{\it Step~3: Proof of~\eqref{78}}. 
Since $\alpha_n\in -J_{t_n}$ and $\delta_{t_n}\to\hat\delta$, the first relation in~\eqref{78} will be established if we show that
\begin{equation} \label{81}
\liminf_{n\to\infty}\alpha_n=\hat\delta. 
\end{equation}
Suppose this is not the case. Then there is $\varepsilon>0$  and a sequence $n_k\to+\infty$ such that $-1\le \alpha_{n_k}\le\hat\delta-\varepsilon$, where the first inequality follows from the fact that $\tilde e_{t_n}'(\alpha_n)\ge0$ and $\tilde e_{t_n}'(-1)\le0$. To simplify notation, we assume that the entire sequence~$\{\alpha_n\}$ satisfies this inequality. It follows that 
\begin{equation} \label{82}
s^+\le \hat s=\frac{1}{t_n}\tilde e_{t_n}'(\alpha_{n})\le\frac{1}{t_n}\tilde e_{t_n}'(\hat\delta-\varepsilon)\quad\mbox{for any $n\ge1$}. 
\end{equation}
Since $\frac{1}{t_n}e_{t_n}(\alpha)$ are convex functions converging to the smooth function~$\tilde e(\alpha)$ for $\alpha\in-\hat J$, by Theorem~25.7 in~\cite{Rock}, we have
$$
\lim_{n\to\infty}\frac{1}{t_n}\tilde e_{t_n}'(\alpha)=\tilde e'(\alpha)\quad\mbox{for any $\alpha\in-\hat J$},
$$
and the limit is uniform on any compact subset of~$-\hat J$. Comparing this with~\eqref{82}, we see that $s^+\le \tilde e'(\hat\delta-\varepsilon)$.
It follows that $\tilde e'$ is constant on the interval~$[\hat\delta-\varepsilon,\hat\delta]$ and, hence, by analyticity and the first relation in~\eqref{SR}, the function~$e(\alpha)$ vanishes. This contradicts the assumption that $\omega_+(\sigma)>0$ and proves~\eqref{81}. 

We now establish the second relation in~\eqref{78}. For any $\gamma\in(0,\hat\delta)$, we have
$$
\tilde e_{t_n}(\alpha_n)=\tilde e_{t_n}(\gamma)+\int_{\gamma}^{\alpha_n}\tilde e_{t_n}'(\alpha)\,\d\alpha
\ge \tilde e_{t_n}(\gamma)+(\alpha_n-\gamma)\tilde e_{t_n}'(0),
$$
where we used the facts that~$\tilde e'$ is nondecreasing and that $\alpha_n>\gamma$ for sufficiently large $n\ge1$, in view of the first relation in~\eqref{78}. It follows that
$$
\liminf_{n\to\infty}\frac{1}{t_n}\tilde e_{t_n}(\alpha_n)\ge \tilde e(\gamma)+(\hat\delta-\gamma)\tilde e'(0). 
$$
Passing to the limit as $\gamma\to\hat\delta$, we obtain the required inequality. 

{\it Step~4: Proof of~\eqref{79}.}
Let us introduce trace class operators 
$$
Q_n=D^{1/2}T_{t_n}D^{1/2},\quad M_n=t_n^{-1}(I-\alpha_nQ_n)^{-1}Q_{n},\quad n\ge1.
$$ 
Since $\alpha_n\in-J_{t_n}$, the operator $I-\alpha_nQ_n$ is strictly positive and, hence, invertible, so that~$M_n$ is well defined. Suppose we have shown that 
\begin{equation} \label{83}
\nu_n\bigr(f(X_n)\bigl)=\mu\bigr(f(Y_n)\bigl), \quad X_n=-t_n^{-1}\ell_{\omega_{t_n}|\omega},
\quad Y_n=\frac12(x,M_nx), \quad n\ge1,
\end{equation}
where $f:\rr\to\rr$ is an arbitrary bounded measurable function and $\mu$ is the centered Gaussian measure on~$\fX$ with the covariance operator~$I$. In this case, taking~$f$ to be the indicator function of the interval~$\cal J_{\hat s,\varepsilon}$, we can write
$$
\nu_n\bigl(B_n(\hat s,\varepsilon)\bigr)=\mu\bigl(\{x\in\fX\,|\,|Y_n(x)-\hat s|<\varepsilon\}\bigr)=:p_n(\varepsilon)\quad\mbox{for any $n\ge1$}.
$$
Thus, the required assertion will be established if we prove that 
\beq
\label{083}
\inf_{n\ge1}p_n(\varepsilon)>0\quad\mbox{for any $\varepsilon>0$}. 
\eeq

To this end, let us assume that we have proved that
\begin{equation} \label{84}
{\mathfrak M}:=\sup_{n\ge 1}\|M_n\|_1<\infty, \quad \tr(M_n)=2\hat s. 
\end{equation}

We now use the following lemma, whose proof is given in the end of this subsection (cf. Lemma~2 in~\cite[Section 3]{BD}.)

\begin{lemma}\label{l4.2} 
Let $\mu$ be the centered Gaussian measure on~$\fX$ with the covariance operator~$I$. Then for any positive numbers~$\kappa$ and~$\varepsilon$ there is $p(\kappa,\varepsilon)>0$ such that
\begin{equation} \label{93}
\mu\bigl(\{x\in\fX\,|\,|(x,Mx)-\tr(M)|<\varepsilon\}\bigr)\ge p(\kappa,\varepsilon)
\end{equation}
for any selfadjoint operator $M\in\cT$ satisfying the inequality $\|M\|_1\le\kappa$. 
\end{lemma}

In view of~\eqref{84}, we have 
$$
Y_n(x)-\hat s=\bigl(x,\tfrac12M_nx\bigr)-\tr\bigl(\tfrac12 M_n\bigr).
$$
Applying Lemma~\ref{l4.2} with $\kappa=2\,{\mathfrak M}$, we see that~\eqref{083} holds. Thus, to complete the proof of the theorem, it remains to establish~\eqref{83} and~\eqref{84}. 

{\it Step~5: Proof of the auxiliary assertions.}
Simple approximation and analyticity arguments show that, to prove~\eqref{83}, is suffices to consider the case in which $f(x)=\e^{\gamma x}$, where $\gamma\in\rr$ is sufficiently small. Thus, we need to check that
\begin{equation} \label{91}
\nu_n\bigl(\exp(-\gamma t_n^{-1}\ell_{\omega_{t_n}|\omega})\bigr)=\mu\bigl(\e^{\gamma Y_n}\bigr). 
\end{equation}
Recalling the construction of~$\alpha_n$ and using the relation $\tilde e_t(\alpha)=-\frac12\log\det(I-\alpha Q_t)$ (see~\eqref{eta}), we write
\begin{align*}
\nu_n\bigl(\exp(-\gamma t_n^{-1}\ell_{\omega_{t_n}|\omega})\bigr)
&=\int_{\fX}\exp\bigl(-(\gamma t_n^{-1}+\alpha_n)\ell_{\omega_{t_n}|\omega}-\tilde e_{t_n}(\alpha_n)\bigr)\omega(\d x)\\ 
&=\exp\bigl(\tilde e_{t_n}(\gamma t_n^{-1}+\alpha_n)-\tilde e_{t_n}(\alpha_n)\bigr)
=\det\bigl(I-\gamma M_n\bigr)^{-1/2}.
\end{align*}
This expression coincides with the right-hand side of~\eqref{91}.

Finally, to prove~\eqref{84}, we first note that the equality follows immediately from the choice of~$\alpha_n$ and the relation $\tilde e_t'(\alpha)=\frac12\tr\bigl((I-\alpha Q_t)^{-1}Q_t\bigr)$. To establish the inequality, we start by using ~\eqref{36} and~\eqref{EQ-UFlow} to
get the bound
\begin{equation} \label{92}
\|Q_n\|_1\le\int_0^{t_n}\|D^{1/2}\varsigma_{-s}D^{1/2}\|_1\d s
\le\frac{M^2}m t_n\|\varsigma\|_1. 
\end{equation}
Writing the spectral decomposition of the compact self-adjoint operator~$M_n$, we easily show that
$$
M_n^-=t_n^{-1}(I+\alpha_nQ_n^-)^{-1}Q_n^-,
$$
where~$A^+$ and~$A^-$ stand the positive and negative parts of a selfadjoint operator~$A$, and we used that fact that $\alpha_n>0$ for sufficiently large~$n$ (see~\eqref{81}). Combining this relation with~\eqref{92}, we derive
$$
\tr(M_n^-)=t_n^{-1}\tr\bigl((I+\alpha_nQ_n^-)^{-1}Q_n^-\bigr)\le \frac{M^2}{m}\|\varsigma\|_1. 
$$
Recalling the second relation in~\eqref{84}, we conclude that
$$
\|M_n\|_1=\tr(|M_n|)=\tr(M_n+2M_n^-)
\le2\left(\hat s+\frac{M^2}{m}\|\varsigma\|_1\right).
$$
The proof of Theorem~\ref{EStheorem-bis} is complete. 
\qed

\medskip
{\it Proof of Lemma~\ref{l4.2}.}
We set $Y(x)=(x,Mx)$ and note that $\mu(Y)=\tr(M)$. Let us denote by~$\{P_I,I\subset\rr\}$ the family of spectral projections for~$M$ and, given a number $\theta>0$, write $M=M^{\le \theta}+M^{>\theta}$, where $M^{\le \theta}=MP_{[-\theta,\theta]}$. 
Accordingly, we represent~$Y$ in the form
$$
Y(x)=Y^{\le \theta}(x)+Y^{>\theta}(x), \quad Y^{\le \theta}(x)=\bigl(x,M^{\le \theta}x\bigr)-\tr\bigl(M^{\le \theta}\bigr).
$$
Now note that  the random variables~$Y^{\le \theta}$ and~$Y^{>\theta}$ are independent under the law~$\mu$. It follows that the probability~$P(M,\varepsilon)$ given by the left-hand side of~\eqref{93} satisfies the inequality
\begin{equation} \label{085}
P(M,\varepsilon)\ge\mu\bigl(\{|Y^{>\theta}|<\varepsilon/2,|Y^{\le\theta}|<\varepsilon/2\}\bigr)
=\mu\bigl(\{|Y^{>\theta}|<\varepsilon/2\}\bigr)\mu\bigl(\{|Y^{\le\theta}|<\varepsilon/2\}\bigr).
\end{equation}
We claim that both factors on the right-hand side of this inequality are separated from zero. Indeed, to estimate the first factor, we note that 
\begin{equation} \label{88}
\kappa\ge \|M\|_1\ge \theta\rank\bigl(M^{>\theta}\bigr),
\end{equation}
where $\rank(M^{>\theta})=:N_\theta$ stands for the rank of~$M^{>\theta}$. 
Denoting by~$\lambda_j$ the eigenvalues of~$M$ indexed in the non-increasing order of their absolute values, we see that
$$
|Y^{>\theta}(x))|=\biggl|\,\sum_{j=1}^{N_\theta}\lambda_j(x_j^2-1)\biggr|\le \kappa\sum_{j=1}^{N_\theta}|x_j^2-1|,
$$
where $\{x_j\}$ are the coordinates of~$x$ in the orthonormal basis formed of the eigenvectors of~$M$.
Combining this with~\eqref{88}, we derive
$$
\mu\bigl\{|Y^{>\theta}(x))|<\varepsilon/2\bigr\}\ge \mu\biggl\{\,\sum_{j=1}^{N_\theta}|x_j^2-1|<\frac{\varepsilon}{2\kappa}\biggr\}
\ge \prod_{j=1}^{N_\theta}\mu\bigl\{|x_j^2-1|<(2\kappa N_\theta)^{-1}\varepsilon\bigr\}
\ge p\bigl(\delta)^{\kappa/\theta},
$$
where $\delta=\varepsilon\theta/(2\kappa^2)$, and $p(\delta)>0$ is the probability of the event $|x^2-1|<\delta$ under the one-dimensional standard normal law. To estimate the second factor in~\eqref{085}, we use the Chebyshev inequality:
\begin{align} \label{90}
\mu\bigl\{|Y^{\le\theta}(x)|<\varepsilon/2\bigr\}&=1-\mu\bigl\{Y^{\le\theta}(x)\ge\varepsilon/2\bigr\}-\mu\bigl\{-Y^{\le\theta}(x)\ge\varepsilon/2\bigr\}
\notag\\
&\ge 1-\mu\bigl(\exp(\gamma Y^{\le\theta}-\gamma\varepsilon/2)\bigr)+\mu\bigl(\exp(-\gamma Y^{\le\theta}-\gamma\varepsilon/2)\bigr),
\end{align}
where $\gamma>0$ is sufficiently small and will be chosen later. We have
\begin{align}
\mu\bigl(\exp(\gamma Y^{\le\theta})\bigr)&=\exp\bigl\{-\gamma\tr\bigl(M^{\le\theta}\bigr)-\tfrac12\log\det\bigl(I-\gamma M^{\le\theta}\bigr)\bigr\} \notag\\
&=\exp\bigl\{-\tfrac{1}{2}\tr\bigl(2\gamma M^{\le\theta}+\log(I-2\gamma M^{\le\theta})\bigr)\bigr\}.\label{091}
\end{align}
Now note that if $4|\gamma|\theta\le 1$, then
$$
2\gamma M^{\le\theta}+\log(I-2\gamma M^{\le\theta})=\sum_{n=2}^\infty\frac{\bigl(-2\gamma M^{\le\theta}\bigr)^n}{n}.
$$
Recalling that $\|M^{\le\theta}\|\le \theta$ and $\|M^{\le\theta}\|_1\le\kappa$ and using the inequality $|\tr(AB)|\le \|A\|_1\|B\|$, it follows that 
$$
\bigl|\tr\bigl(2\gamma M^{\le\theta}+\log(I-2\gamma M^{\le\theta})\bigr)\bigr|\le \sum_{n=2}^\infty|2\gamma\theta|^{n-1}2|\gamma|\kappa
\le 8\kappa\gamma^2\theta.
$$
Substituting this into~\eqref{091}, we see that, if $|\gamma|\le(4\theta)^{-1}$, then $\mu\bigl(\exp(\gamma Y^{\le\theta})\bigr)\le \exp\bigl(4\kappa\gamma^2\theta\bigr)$. A similar estimate holds for $\mu\bigl(\exp(-\gamma Y^{\le\theta})\bigr)$. Combining  these inequalities  with~\eqref{90} and choosing $\gamma=\frac{\varepsilon}{16\kappa\theta}$, we derive
$$
\mu\bigl\{|Y^{\le\theta}(x)|<\varepsilon/2\bigr\}\ge 1-2\exp\bigl(4\kappa\gamma^2\theta-\gamma\varepsilon/2\bigr)
=1-2\exp\bigl(-\tfrac{\varepsilon^2}{64\kappa\theta}\bigr).
$$
The right-hand side of this inequality can be made greater than zero by choosing a sufficiently small~$\theta>0$ which will depend only on~$\kappa$ and~$\varepsilon$. 
\qed

\subsection{Proof of Theorem \ref{GCtheorem}}
The proof of this result is verty similar to that of Theorems~\ref{EStheorem} and~\ref{EStheorem-bis}, and we shall only outline the proof. 

\noindent{\bf Part (1)} Follows from H\"older's inequality as in the proof of Proposition \ref{analprop} (2).

\noindent{\bf Part (2)}
Since $0\in J_t^+$, the fact that~$J_t^+$ is an interval follows immediately from the following property: if $\alpha\in J_t^+$, then $\theta\alpha\in J_t^+$ for $\theta\in(0,1)$. To prove the analyticity, note that, by Eq.~\eqref{gausspert}, one has 
$$
\e^{-\alpha\ell_{\omega_t|\omega}}\d\omega_+
=\sqrt{\frac{\bigl(\det(I+DT_t)\bigr)^{-\alpha}}{\det(I-\alpha D_+ T_t)}}\,\d\omega_{(D_+^{-1}-\alpha T_t)^{-1}}.
$$
This relation implies that the function
\begin{align}
e_{t+}(\alpha)
&=-\frac\alpha2\log\det(I+D_+T_t)-\frac12\log\det(I-\alpha D_+T_t)
\label{EQ-etplus}\\
&=-\frac\alpha2\log\det(I+D^{1/2}T_tD^{1/2})-\frac12\log\det(I-\alpha D_+^{1/2}T_tD_+^{1/2})\nonumber
\end{align}
is real analytic in~$\alpha$ on the open interval  defined by the 
condition $I-\alpha D_+^{1/2}T_tD_+^{1/2}>0$ and takes the value~$+\infty$ on its complement (where the intersection of the spectrum of $I-\alpha D_+^{1/2}T_tD_+^{1/2}$ with the negative half-line is nonempty). The above inequality coincides with the one defining~$J_t^+$. 

\noindent{\bf Part (3)} 
The fact that $\ubar J^+$ is an interval follows immediately from its definition. 
To prove that $J_t^+\supset (-\delta,\delta)$, note that, in view of Hypothesis \Gtwo{}, for any $t,\alpha\in\rr$ we have
$$
I-\alpha D_+^{1/2}T_tD_+^{1/2}=D_+^{1/2}(D_+^{-1}-\alpha(D_t^{-1}-D^{-1}))D_+^{1/2}\ge\frac{\delta-|\alpha|}{\delta+1}.
$$
This expression is positive for $|\alpha|<\delta$.

To prove the existence of limit~\eqref{proto-grf-gc} and its analyticity on~$\ubar J^+$, we repeat the argument used in the proof of Theorem~\ref{EStheorem-bis} (2). Namely, let us introduce the family of operators $D_t^+(\alpha)=(D_+^{-1}-\alpha T_t)^{-1}$, which are well defined for $\alpha\in(-\delta,\delta)$. Then the following analogue of relation~\eqref{69} is valid:
$$
\omega_{D_t^+(\alpha)}(f)
=\frac{\omega(\e^{-\alpha\ell_{\omega_t|\omega}}f)}{\omega(\e^{-\alpha\ell_{\omega_t|\omega}})}
\quad\mbox{for $f\in L^1(\fX,\d\omega_{D_t^+(\alpha)})$}.
$$
The argument used in the derivation of~\eqref{e-form} gives that  
$$
\frac1t e_{t+}(\alpha)
=-\int_0^\alpha\int_0^1\omega_{{D}_t^+(\gamma)}(\sigma_{-ts})\,\d s\d\gamma,
$$
while Hypothesis \Gtwo{} and the relation $\e^{r\cL}D_+\e^{r\cL^*}=D_+$ valid for $r\in\rr$ imply that
$$
\e^{-ts\cL}{D}_t^+(\gamma)\e^{-ts\cL^\ast}
=\bigl(D_+^{-1}-\gamma(D_{t(1-s)}^{-1}-D_{-ts}^{-1})\bigr)^{-1}
\le M\left(1-\frac{|\gamma|}{\delta}\right)^{-1}.
$$
Following again the argument in the proof of Theorem \ref{EStheorem-bis}~(2), for $\alpha\in(-\delta,\delta)$ we derive
\begin{equation} \label{73}
e_+(\alpha)=\lim_{t\to\infty}\frac1te_{t+}(\alpha)
=-\int_0^\alpha\omega_{\overline{D}_{1-\gamma}}(\sigma)\,\d\gamma.
\end{equation}
Now note that $\overline{D}_{1-\gamma}=\vartheta\overline{D}_{\gamma}\vartheta$, whence it follows $\omega_{\overline{D}_{1-\gamma}}(\sigma)=\omega_{\overline{D}_{\gamma}}(\sigma\circ\vartheta)=-\omega_{\overline{D}_{\gamma}}(\sigma)$. Substituting this into~\eqref{73} and recalling~\eqref{neckpain}, we see that 
\begin{equation} \label{74}
e_+(\alpha)=\int_0^\alpha\omega_{\overline{D}_{\gamma}}(\sigma)\,\d\gamma=e(\alpha)\quad\mbox{for $\alpha\in(-\delta,\delta)$}. 
\end{equation}
We have thus established the existence of limit~\eqref{proto-grf-gc} on the interval $(-\delta,\delta)\subset \ubar J^+$. The fact that it exists for any $\alpha\in\ubar J^+$ and defines a real-analytic function can be proved with the help of Vitali's theorem (cf.\ proof of Part~(2) of Theorem~\ref{EStheorem-bis}). Finally, relation~\eqref{33} is established by the same argument as~\eqref{028}.

\noindent{\bf Parts (4--6)} The proofs of the large deviation principle, central limit theorem, and strong law of large numbers for the time average of the entropy production functional under the limiting law~$\omega_+$ are exactly the same as for~$\omega$ (see Parts (3--5) of Theorem \ref{EStheorem}), and therefore we will omit them.

\noindent{\bf Parts (7)} 
The fact that the functions $e_+(\alpha)$ and $e(\alpha)$ coincides on the intersection $\ubar J^+\cap \ubar J$ follows from~\eqref{74} and their analyticity. The equality of the corresponding rate functions on a small interval around ~$\omega_+(\sigma)$ is a straightforward consequence of~\eqref{74} and the definition of the Legendre transform.
\qed

\end{document}
%
%

\subsubsection*{Non-converging intervals}
Let us assume that the generator~$\cL$ and a unit vector $\varphi\in\cK$ are such that $\|\e^{t\cL}\|\le C$ for all $t\in\rr$, $\varphi_t=\e^{-t\cL^*}\varphi\to0$ weakly in~$\cK$ as $t\to+\infty$, and 
$$
\label{203}
0<A_-:=\liminf_{t\to+\infty}\|\varphi_t\| <\limsup_{t\to+\infty}\|\varphi_t\|=:A_+.
$$
Let the initial state be a centered Gaussian measure whose covariance operator is given by (cf.~\eqref{200})
$$
D=\bigl((D^0)^{-1}+\lambda P_\varphi\bigr)^{-1},
$$
where $\lambda>0$ is a parameter and $D^0:\cK\to\cK$ is a positive operator such that $D^0=\e^{t \cL}D^0\e^{t \cL^*}$ for all $t\in\rr$.  A more realistic example for which these conditions are satisfied is given in Section~\ref{s3.3}. 

We have $D_t^{-1}=(D^0)^{-1}+\lambda P_{\varphi_t}$ for $t\in\rr$. Using this relation and repeating the above arguments, it is not difficult to check that Hypotheses~\Gone--\Gthree{} are satisfied with $D_\pm=D^0$. Let us choose two increasing sequences $\{t_k^\pm\}$ going to~$+\infty$ as $k\to\infty$ such that 
\begin{equation} \label{206}
\lim_{k\to\infty}\|\varphi_{t_k^\pm}\|=A_\pm.
\end{equation}
We now derive an asymptotic formula for the intervals~$J_t=(-\delta_t,1+\delta_t)$ with $t=t_k^\pm$ as $\lambda\to+\infty$. Since~$J_t$ is determined by the condition of positivity of the self-adjoint operator (cf.~\eqref{204})
$$
D^{-1} +\alpha(D_t^{-1} - D^{-1})=\lambda C_t(\lambda^{-1},\alpha), \quad 
C_t(\lambda^{-1},\alpha)=\lambda^{-1}(D^0)^{-1}+(1-\alpha) P_\varphi+\alpha P_{\varphi_t},
$$
to find~$\delta_t$, it suffices to determine the lower bound of the spectrum of $C_t(\lambda,\alpha)$. The weak convergence $\varphi_t\to0$ as $t\to\infty$ implies that the vectors~$\varphi$ and~$\varphi_t$ are linearly independent for~$t\gg1$, and a simple calculation shows that $\sigma(C_t(0,\alpha))=\{0,\mu_t^\pm(\alpha)\}$ for $t\gg1$, where $\mu_t^\pm(\alpha)$ are the solutions of the quadratic equation 
$$
\mu^2-(1-\alpha+\alpha\|\varphi_t\|^2)\mu+\alpha(1-\alpha)\bigl(\|\varphi_t\|^2-(\varphi,\varphi_t)^2\bigr)=0.
$$ 

begin{thebibliography}{999999}

\bibitem[AJPP]{AJPP} Aschbacher, W., Jak\v si\'c, V., Pautrat, Y., and Pillet, C.-A.:
\newblock  Transport properties of quasi-free fermions.
\newblock  J. Math. Phys. {\bf 48}, 032101-1--28 (2007).

\bibitem[Ba1]{Ba1} Baladi, V.:
\newblock {\em Positive Transfer Operators and Decay of Correlations}.
\newblock Advanced Series in Nonlinear Dynamics {\bf 16}.
World Scientific, River Edge, NJ (2000).

\bibitem[Ba2]{Ba2} Baladi, V., and Tsujii, M.:
\newblock Anisotropic H\"older and Sobolev spaces for hyperbolic diffeomorphisms.   
\newblock Ann. Inst. Fourier,  {\bf 57},  127--154 (2007).

\bibitem[Ba3]{Ba3} Baladi, V., and Tsujii, M.:  
\newblock  Dynamical determinants and spectrum for hyperbolic diffeomorphisms.  
\newblock  In{ \em  Probabilistic and Geometric Structures in Dynamics}.  
K. Burns, D. Dolgopyat and Ya. Pesin (editors). Contemp. Math. {\bf 469}, 29--68
(2008).


\bibitem[BGM]{BGM} Bonetto, F.,  Gentile, G., and  Mastropietro, V.:  
\newblock Electric fields on a surface of constant negative curvature. 
\newblock Erg. Th. Dyn. Sys. {\bf  20}, 681--696 (2000).

\bibitem[BK]{BK}Brin, M., and  Katok, A.:
\newblock  On local entropy.
\newblock Lecture Notes in Mathematics {\bf 1007}, 30--38. Springer, Berlin, (1983). 

\bibitem[BKL]{BKL} Blank, M., Keller, G., and  Liverani, C.:
\newblock Ruelle-Perron-Frobenius spectrum for Anosov maps.
\newblock Nonlinearity {\bf 15}, 1905--1973 (2002).

\bibitem[BL]{BL} Butterley, O., and Liverani, C.:
\newblock Smooth Anosov flows: correlation spectra and stability. 
\newblock Journal of Modern Dynamics {\bf  1},  301--322  (2007).

\bibitem[Bo1]{Bo1} Bowen, R.:  
\newblock Some systems with unique equilibrium state. 
\newblock Math. Systems Theory {\bf 8}, 193--202 (1974).

\bibitem[Bo2]{Bo2} Bowen, R.:
\newblock {\em Equilibrium States and the Ergodic Theory of Anosov
Diffeomorphisms.}
\newblock Lecture Notes in Mathematics {\bf 470}. Springer,
Berlin (1975).

\bibitem[BR]{BR} Bratteli, O., and Robinson, D. W.: 
\newblock {\em Operator Algebras and Quantum Statistical Mechanics 1.}
\newblock  Springer, Berlin (1987).


\bibitem[BS]{BS} Beck, C., and Schlo\"ogl.: 
\newblock{Thermodynamics of chaotic systems}.
\newblock Cambridge Nonlinear Science Series {\bf 4}. Cambridge University Press, 
Cambridge (1993).


\bibitem[CELS1]{CELS1}Chernov, N. I., Eyink, G.L., Lebowitz, J.L., and Sinai, 
Ya.G.:
\newblock Derivation of Ohm's law in a deterministic mechanical model.
\newblock Phys. Rev. Lett. {\bf 70}, 2209--2212 (1993).

\bibitem[CELS2]{CELS2} Chernov, N. I., Eyink, G.L., Lebowitz, J.L., and Sinai, 
Ya.G.:
\newblock Steady-state electrical conduction in the periodic Lorentz gas. 
\newblock Commun. Math. Phys. {\bf 154}, 569--601 (1993).

\bibitem[CG]{CG}  Cohen, E. G. D., and  Gallavotti, G.:
\newblock Note on two theorems in nonequilibrium statistical mechanics.
\newblock J. Stat. Phys. {\bf 96}, 1343--1349 (1999).

\bibitem[Ch1]{Ch1} Chernov, N. I.: 
\newblock Sinai billiards under small external forces. 
\newblock Ann. Henri Poincar\'e {\bf 2}, 197--236 (2001).

\bibitem[Ch2]{Ch2} Chernov, N. I.: 
\newblock Sinai billiards under small external forces 2. 
\newblock Ann. Henri Poincar\'e {\bf 9}, 91--107 (2008).

\bibitem[CL]{CL} Chernov, N. I., and  Lebowitz, J. L.: 
\newblock Stationary nonequilibrium states in boundary-driven Hamiltonian 
systems: Shear flow. 
\newblock J. Stat. Phys.  {\bf 86},  953--990 (1997). 

\bibitem[CWW]{CWW} Carberry, D.M.,  Williams, S.R.,  Wang, G.M.,   Sevick, E.M., and  Evans J. D.:
\newblock The Kawasaki identity and the fluctuation Theorem.
\newblock J. Chem. Phys. {\bf 121}, 8179--8182 (2004).



\bibitem[DDM]{DDM}Derezi\'nski, J., De Roeck, W., and Maes, C.:
\newblock Fluctuations of quantum currents and unravelings
of master equations.
\newblock J. Stat. Phys. {\bf 131}, 341--356 (2008).

\bibitem[DGT]{DGT} Dorfman, J.R, Gilbert T., Tasaki, S.: 
\newblock An analytical construction of the SRB measures for Baker-type maps.
\newblock Chaos,  {\bf 8}, 424--442 (1998).

\bibitem[Do]{Do} Dorfman, J.R.: 
\newblock {\em An Introduction to Chaos in Nonequilibrium Statistical Mechanics.} 
\newblock Cambridge University Press,  Cambridge (1999).

\bibitem[Do1]{Do1}Dolgopyat, D.: 
\newblock Decay of correlations in Anosov flows. 
\newblock Ann. Math. {\bf 147}, 357--390 (1998).

\bibitem[Do2]{Do2}Dolgopyat, D.: 
\newblock Prevalence of rapid mixing in hyperbolic flows.
\newblock Erg. Th. Dyn. Sys. {\bf 18}, 1097--1114 (1998).

\bibitem[DZ]{DZ} Dembo, A., and Zeitouni, O.: 
\newblock {\em Large Deviations Techniques and Applications.} 
\newblock Second edition. Applications of Mathematics, {\bf 38}. 
Springer, New York (1998). 



\bibitem[ECM]{ECM}  Evans, D.J., Cohen, E.G.D., and Morriss, G.P.:
\newblock Probability of second law violation in shearing steady flows. 
\newblock Phys. Rev. Lett. {\bf 71}, 2401--2404 (1993).

\bibitem[EH1]{EH1} Eckmann, J.-P., and  Hairer, M.:
\newblock Non-equilibrium statistical mechanics of strongly anharmonic 
chains of oscillators.
\newblock Commun. Math. Phys. {\bf  212}, 105--164 (2000).

\bibitem[EH2]{EH2} Eckmann, J.-P., and Hairer, M.:
\newblock Spectral properties of hypoelliptic operators.
\newblock Commun. Math. Phys.  {\bf 235}, 233--253 (2003).

\bibitem[El]{El} Ellis, R.S.:
\newblock{\em Entropy, Large Deviations, and Statistical Mechanics.}
\newblock Springer, Berlin (1985). Reprinted in the series Classics of Mathematics
(2006).

\bibitem[EM]{EM} Evans, D.J., and Morriss, G.P.:
\newblock {\em Statistical Mechanics of Nonequilibrium Fluids.} 
\newblock Academic Press, New York (1990).  

\bibitem[EPR1]{EPR1} Eckmann, J.-P., Pillet, C.-A., and Rey-Bellet, L.:
\newblock Non-equilibrium statistical mechanics of anharmonic chains
coupled to two heat baths at different temperatures.
\newblock Commun. Math. Phys.  {\bf 201}, 657--697 (1999).

\bibitem[EPR2]{EPR2} Eckmann, J.-P., Pillet, C.-A., and Rey-Bellet, L.:
\newblock Entropy production in nonlinear, thermally driven Hamiltonian 
systems.
\newblock J. Stat. Phys.  {\bf 95}, 305--331 (1999).

\bibitem[ES]{ES} Evans, D.J., and Searles, D.J.: 
\newblock Equilibrium microstates which generate second law violating steady 
states. 
\newblock Phys Rev. E {\bf 50}, 1645--1648 (1994).


\bibitem[Ga1]{Ga1} Gallavotti, G.:
\newblock Chaotic hypothesis: Onsager reciprocity and fluctuation-dissipation
theorem.
\newblock J. Stat. Phys.  {\bf 84}, 899--925 (1996).

\bibitem[Ga2]{Ga2} Gallavotti, G.:
\newblock Dynamical ensembles equivalence in fluid mechanics.
\newblock Physica D  {\bf 105}, 163--184 (1997).

\bibitem[Ga3]{Ga3} Gallavotti, G.:
\newblock Chaotic principle: some applications to developed turbulence.
\newblock J. Stat. Phys. {\bf 86}, 907--934 (1997).

 

\bibitem[GC1]{GC1} Gallavotti, G., and  Cohen, E.G.D.:
\newblock Dynamical ensembles in nonequilibrium statistical mechanics.
\newblock Phys. Rev. Lett. {\bf  74}, 2694--2697 (1995).

\bibitem[GC2]{GC2} Gallavotti, G., and  Cohen, E. G. D.:
\newblock Dynamical ensembles in stationary states.
\newblock J. Stat. Phys. {\bf  80}, 931--970  (1995).

\bibitem[Ge]{Ge} Gentile, G.:
\newblock Large deviation rule for Anosov flows.
\newblock Forum Math. {\bf 10}, 89--118 (1998).

\bibitem[GL1]{GL1} Gouezel, S., and Liverani, C.:
\newblock  Banach spaces adapted to Anosov systems
\newblock Erg. Th.  Dyn.  Sys. {\bf 26}, 189--217 (2006).

\bibitem[GL2]{GL2} Gouezel, S., and Liverani, C.:
\newblock Compact locally maximal hyperbolic sets for smooth maps: 
fine statistical properties.
\newblock  J. Diff. Geom. {\bf 79}, 433--477 (2008).

\bibitem[GM]{GM} de Groot, S.R., and Mazur, P.:
\newblock{\em Nonequilibrium Thermodynamics.}
\newblock NorthHolland,  Amsterdam (1962).

\bibitem[GrRy]{GrRy} Gradshteyn, I.S. and Ryzhik, I.M.:
\newblock {\em Tables of Integrals, Series and Products.} 7th edition.
\newblock  Academic Press, Amsterdam (2007).

\bibitem[GR]{GR} Gallavotti, G., and Ruelle, D.:
\newblock SRB states and nonequilibrium statistical mechanics close to equilibrium.
\newblock  Commun. Math. Phys. {\bf 190}, 279--285 (1997). 

\bibitem[Gr1]{Gr1} Green, M.S.:
\newblock Markoff random processes and the statistical mechanics of time-dependent phenomena.
\newblock J. Chem. Phys. {\bf 20}, 1281--1295 (1952).

\bibitem[Gr2]{Gr2} Green, M.S.:
\newblock Markoff random processes and the statistical mechanics of time-dependent phenomena.
II. Irreversible processes in fluids. 
\newblock J. Chem. Phys. {\bf  22}, 398--413 (1954).

\bibitem[GRS]{GRS} Gallavotti, G., Rondoni, L., and Segre E.:
\newblock Lyapunov spectra and nonequilibrium ensembles equivalence
in 2D fluid mechanics.
\newblock Physica D {\bf 187}, 338--357 (2004).

\bibitem[HH]{HH}  Hennion, H., and Herv\'e, L.:
\newblock  {\em Limit Theorems for Markov Chains and Stochastic Properties of 
Dynamical Systems by Quasi-Compactness.}
\newblock  Lecture Notes in Mathematics {\bf 1766}. Springer, Berlin (2001).

\bibitem[HHP]{HHP} Holian, B.L., Hoover, W.G., and Posch, H.A.: 
\newblock Resolution of Loschmidt's paradox: The origin of irreversible 
behavior in reversible atomistic dynamics. 
\newblock Phys. Rev. Lett. {\bf 59}, 10--13 (1987).

\bibitem[Ho]{Ho} Hoover, E.G.: 
\newblock {\em Molecular Dynamics}.
\newblock Lecture Notes in Physics {\bf 258}. Springer, Berlin (1986).


\bibitem[Je]{Je} Jenkinson, O.: 
\newblock Rotation, entropy, and equilibrium states.
\newblock Trans. AMS {\bf 353}, 3713--3739 (2001).

\bibitem[JLTP]{JLTP} Jak${\check {\rm s}}$i\'c  V., Larochelle, V., Tomberg, A., and  
Pillet, C.-A.:
\newblock In preparation.

\bibitem[JP1]{JP1} Jak\v si\'c, V., and Pillet, C.-A.:
\newblock On entropy production in quantum statistical mechanics. 
\newblock Commun. Math. Phys. {\bf 217},  285--293 (2001).

\bibitem[JP2]{JP2} Jak\v si\'c, V., and Pillet, C.-A.:
\newblock Mathematical theory of non-equilibrium quantum statistical mechanics.
\newblock  J. Stat. Phys. {\bf 108}, 787--829 (2002).

\bibitem[JP3]{JP3} Jak\v si\'c, V., and Pillet, C.-A.:
\newblock  Non-equilibrium steady states of finite quantum systems coupled to 
thermal reservoirs.
\newblock  Commun. Math. Phys. {\bf 226}, 131--162 (2002).

\bibitem[JPP]{JPP} Jak\v si\'c, V., Pautrat, Y., and Pillet, C.-A.:
\newblock A quantum central limit theorem for sums of IID random variables.
\newblock J. Math. Phys. {\bf 51}, 015208 (2010).


\bibitem[Ka]{Ka} van Kampen, N.:
\newblock The case against linear response theory. 
\newblock Physica Norvegica {\bf 5}, 279--284 (1971).

\bibitem[KH]{KH} Katok, A., and Hasselblatt, B.:
\newblock {\em Introduction to the Modern Theory of Dynamical Systems.}
\newblock Encyclopedia of mathematics and its applications {\bf 54}.
Cambridge University Press, Cambridge (1995). 


\bibitem[Ki]{Ki} Kifer, Y.:
\newblock Large deviations, averaging and periodic orbits of dynamical systems.
\newblock Commun. Math. Phys. {\bf 162}, 33--46 (1994).

\bibitem[KKPW]{KKPW} Katok, A.,  Knieper, G.,   Pollicott, M., and Weiss, H.:
\newblock Differentiability of entropy for Anosov and geodesic flows.
\newblock  Bulletin AMS {\bf 22}, 285--293 (1990).

\bibitem[Ko]{Ko} Korevaar, J.: 
\newblock {\em Tauberian Theory. A Century of Developments.} 
\newblock Springer, Berlin, (2004).

\bibitem[Ku1]{Ku1} Kurchan, J.: 
\newblock Fluctuation theorem for stochastic dynamics.  
\newblock J. Phys. A  {\bf 31}, 3719--3729 (1998).

\bibitem[Ku2]{Ku2} Kurchan, J.: 
\newblock A quantum fluctuation theorem.  
\newblock Arxiv preprint cond-mat/0007360 (2000).
 
\bibitem[Kub]{Kub} Kubo, R.: Statistical-mechanical theory of irreversible processes I. 
General theory and simple applications to magnetic and conduction problems.
\newblock J. Phys. Soc. Jap. {\bf 12}, 570--586 (1957).

\bibitem[KTH]{KTH} Kubo, R.,  Toda, M., and Hashitsume, N.:
\newblock {\em  Statistical Physics II. Nonequilibrium Statistical Mechanics.}
\newblock Second edition. Springer Series in Solid-State Sciences {\bf 31}.
Springer, Berlin (1991).


\bibitem[Li1]{Li1} Liverani, C.:
\newblock On contact Anosov flows.
\newblock Ann. of Math. {\bf 159},  1275--1312 (2004).

\bibitem[Li2]{Li2} Liverani, C.:
\newblock Fredholm determinants, Anosov maps and Ruelle resonances.
\newblock Discrete and Continuous Dynamical Systems {\bf 13},
1203--1215 (2005). 

\bibitem[Li3]{Li3} Liverani, C.:
\newblock Private communication. 

\bibitem[LS2]{LS2} Lebowitz, J. L., and Spohn, H.: 
\newblock A Gallavotti--Cohen-type symmetry in the large deviation functional 
for  stochastic dynamics.  
\newblock J. Stat. Phys. {\bf 95}, 333--365  (1999).

\bibitem[LT]{LT}Liverani, C., and Tsujii, M.: 
\newblock Zeta functions and dynamical systems
\newblock Nonlinearity {\bf  19},  2467--2473 (2006).


\bibitem[Ma1]{Ma1} Maes, C.:
\newblock The fluctuation theorem as a Gibbs property.
\newblock  J. Stat. Phys. {\bf 95}, 367--392 (1999).

\bibitem[Ma2]{Ma2}  Maes, C: 
\newblock On the origin and the use  of fluctuation relations for the entropy.
\newblock S\'eminaire Poincar\'e {\bf 2}, 29--62  (2003).

\bibitem[MD]{MD} Morriss, G.P. and Dettmann, C.P.: 
\newblock Thermostats: Analysis and application.
\newblock Chaos {\bf 8}, 321--336 (1998)

\bibitem[MN]{MN} Maes, C., and Neto\v cn\'y, K.:
\newblock Time-reversal and entropy. 
\newblock J. Stat. Phys.  {\bf 110},  269--310 (2003).


\bibitem[MRV]{MRV} Maes, C.,  Redig, F., and  Verschuere, M.: 
\newblock From global to local fluctuation theorems. 
\newblock Mosc. Math. J.  {\bf 1}, 421--438 (2001).

\bibitem[MV]{MV} Maes, C., and Verbitskiy, E.:
\newblock Large deviations and a fluctuation symmetry for chaotic
homeomorphisms.
\newblock Commun. Math. Phys.  {\bf 233}, 137--151 (2003).


\bibitem[On1]{On1} Onsager, L.:
\newblock Reciprocal relations in irreversible processes I.
\newblock Phys. Rev. {\bf 37}, 405--426 (1931).

\bibitem[On2]{On2} Onsager, L.:
\newblock Reciprocal relations in irreversible processes II.
\newblock Phys. Rev. {\bf 38}, 2265--2279 (1931).

\bibitem[OP]{OP} Ohya, M.  and  Petz, D.: 
\newblock {\em  Quantum Entropy and its Use.} Second edition.
\newblock Springer,  Berlin (2004).


\bibitem[PH]{PH} Posch, H.A., and Hoover, W.G.: 
\newblock Nonequilibrium molecular dynamics of a classical fluid. 
\newblock In {\em Molecular liquids, new perspectives in physics and chemistry,}
527--547. Kluwer (1992).

\bibitem[Pi]{Pi} Pillet, C.-A.:
\newblock Entropy production in classical and quantum systems.
\newblock Markov Proc. Related Fields {\bf 7},  145--157 (2001).

\bibitem[Po]{Po}  Pollicott, M.:
\newblock Stability of mixing rates for axiom A attractors. 
\newblock Nonlinearity {\bf 16}, 567--578 (2003).

\bibitem[PS1]{PS1} Pugh, C., and Shub, M.:
\newblock Stable manifolds and hyperbolic sets.
\newblock In {\em Proc. Symp. Pure Math. Vol XIV, Berkeley, Calif. 1968,} 133--163. 
Amer. Math. Soc., Providence, 1970.


\bibitem[Re]{Re} R\'enyi, A.:  
\newblock On measures of information and entropy.
\newblock In {\em Proc. 4th Berkeley Sympos. Math. Statist. and Prob., Vol. I,}
547--561. Univ. California Press, Berkeley, Calif. (1961). 


\bibitem [Ro]{Ro} de Roeck, W.:
\newblock Large deviation generating function  for currents in the Pauli-Fierz model. 
\newblock Rev. Math. Phys. {\bf 21}, 549-585 (2009).

\bibitem[RT1]{RT1} Rey-Bellet, L., and Thomas, L.E.:
\newblock Exponential convergence to non-equilibrium stationary states in
classical statistical mechanics.
\newblock Commun. Math. Phys. {\bf 225}, 305--329 (2002).

\bibitem[RT2]{RT2} Rey-Bellet, L., and Thomas, L. E.:
\newblock Fluctuations of the entropy production in anharmonic chains.
\newblock Ann. Henri Poincar\'e  {\bf 3}, 483--502 (2002).


\bibitem[Ru1]{Ru1} Ruelle, D.:
\newblock {\em Thermodynamic Formalism. 
The Mathematical Structure of Equilibrium Statistical Mechanics.}
\newblock Second edition. Cambridge University Press, Cambridge (2004).

\bibitem[Ru2]{Ru2} Ruelle, D.:
\newblock Entropy production in nonequilibrium statistical mechanics. 
\newblock Commun. Math. Phys. {\bf 189}, 365--371 (1997).

\bibitem[Ru3]{Ru3} Ruelle, D.: 
\newblock Differentiation of SRB states. 
\newblock Commun. Math. Phys.  {\bf 187},  227--241 (1997).
\newblock Correction and complements.
\newblock Commun. Math. Phys. {\bf 234}, 185--190 (2003).

\bibitem[Ru4]{Ru4} Ruelle, D.: 
\newblock Smooth dynamics and new theoretical ideas in nonequilibrium statistical 
mechanics. 
\newblock J. Stat. Phys. {\bf 95}, 393--468 (1999).

\bibitem[Ru5]{Ru5} Ruelle, D.: 
\newblock A remark on the equivalence of isokinetic and 
isoenergetic thermostats in the thermodynamic limit.
\newblock  J. Stat. Phys. {\bf 100}, 757--763 (2000). 

\bibitem[Ru6]{Ru6} Ruelle, D.: 
\newblock Positivity of entropy production in nonequilibrium statistical mechanics. 
\newblock J. Stat. Phys. {\bf 85}, 1--23 (1996).

\bibitem[RY]{RY} Rey-Bellet, L., and Young, L.-S.: 
\newblock Large deviations in non-uniformly hyperbolic dynamical systems.
\newblock Erg. Th. Dyn. Sys. {\bf 28}, 587--612 (2008).

\bibitem[Sh]{Sh} Shapero, D.: 
\newblock Student project, unpublished.

\bibitem[Si]{Si}  Sinai, Y.:
\newblock Gibbs measures in ergodic theory.
\newblock  Uspehi Mat. Nauk {\bf 27}, 21--64 (1972). 
English translation: Russian Math. Surveys {\bf 27}, 21--69 (1972).

\bibitem[TG]{TG} Tasaki, S., and Gaspard, P.:
\newblock Fick's law and fractality of nonequilibrium
stationary states in a reversible multibaker map.
\newblock J. Stat. Phys. {\bf  81}, 935--987 (1995).

\bibitem[TM]{TM} Tasaki, S., and Matsui, T.: 
\newblock Fluctuation theorem, non-equilibrium steady states and 
Maclennan-Zubarev ensembles  of a class of large systems. 
\newblock Fundamental Aspects of Quantum
Physics, Tokyo (2001); QP-PQ: Quantum Probab. White Noise Anal., 17, 100 (World Sci.,
River Edge NJ 2003).



\bibitem[Wa1]{Wa1} Walters, P.:
\newblock {\em An Introduction to Ergodic Theory.}
\newblock Graduate Texts in Mathematics  {\bf 79}. Springer, Berlin (1982).


\bibitem[WL]{WL} Wojtkowski, M.P., and Liverani, C.:
\newblock Conformally symplectic dynamics and symmetry of the Lyapunov spectrum.
\newblock Commun. Math. Phys. {\bf 194}, 47--60 (1998).


\bibitem[Yo]{Yo} Young, L.-S.: 
\newblock Statistical properties of dynamical systems with some hyperbolicity.
\newblock  Ann. Math. {\bf 147}, 585--650 (1998).





\end{thebibliography}

\end{document}